\documentclass[10pt,journal,compsoc]{IEEEtran/IEEEtran}

\pdfoutput=1

\ifCLASSOPTIONcompsoc
  \usepackage[nocompress]{cite}
\else
  \usepackage{cite}
\fi

\usepackage{mathptmx}
\usepackage{graphicx}
\usepackage{times}
\usepackage{xspace}
\usepackage{verbatim}
\usepackage{multirow}
\usepackage{bigstrut}
\usepackage{tabularx}
\usepackage{booktabs}
\usepackage{adjustbox}
\usepackage{subfigure}
\usepackage{xcolor}
\usepackage{amssymb}
\usepackage{pifont}

\usepackage{color,colortbl}
\let\subparagraph\relax
\usepackage{titlesec}
\titleformat{\paragraph}
{\normalfont\normalsize\bfseries}{\theparagraph}{.5em}{}
\titlespacing*{\paragraph}
{0pt}{1.25ex plus 1ex minus .2ex}{1ex plus .1ex}

\definecolor{spruce}{RGB}{254,109,23}
\newcommand{\classspruce}{\crule[spruce]{0.25cm}{0.25cm}\xspace}
\definecolor{lodgepole}{RGB}{92,186,86}
\newcommand{\classlodgepole}{\crule[lodgepole]{0.25cm}{0.25cm}\xspace}
\definecolor{ponderosa}{RGB}{220,68,87}
\newcommand{\classponderosa}{\crule[ponderosa]{0.25cm}{0.25cm}\xspace}
\definecolor{cottonwood}{RGB}{166,118,194}
\newcommand{\classcottonwood}{\crule[cottonwood]{0.25cm}{0.25cm}\xspace}
\definecolor{aspen}{RGB}{133,75,70}

\definecolor{douglas}{RGB}{217,106,188}
\newcommand{\classdouglas}{\crule[douglas]{0.25cm}{0.25cm}\xspace}
\definecolor{krumm}{RGB}{56,113,193}
\newcommand{\classkrumm}{\crule[krumm]{0.25cm}{0.25cm}\xspace}
\newcommand\crule[3][black]{\textcolor{#1}{\rule{#2}{#3}}}

\definecolor{classred}{RGB}{214,39,40}
\newcommand{\classred}{\crule[classred]{0.25cm}{0.25cm}\xspace}
\definecolor{classblue}{RGB}{31,119,180}
\newcommand{\classblue}{\crule[classblue]{0.25cm}{0.25cm}\xspace}
\definecolor{classorange}{RGB}{255,127,14}
\newcommand{\classorange}{\crule[classorange]{0.25cm}{0.25cm}\xspace}
\definecolor{classgreen}{RGB}{44,160,44}
\newcommand{\classgreen}{\crule[classgreen]{0.25cm}{0.25cm}\xspace}

\definecolor{tr}{RGB}{188,189,34}
\newcommand{\classtr}{\crule[tr]{0.25cm}{0.25cm}\xspace}
\definecolor{tc}{RGB}{127,127,127}
\newcommand{\classtc}{\crule[tc]{0.25cm}{0.25cm}\xspace}
\definecolor{co}{RGB}{31,119,180}
\newcommand{\classco}{\crule[co]{0.25cm}{0.25cm}\xspace}
\definecolor{hi}{RGB}{214,39,40}
\newcommand{\classhi}{\crule[hi]{0.25cm}{0.25cm}\xspace}
\definecolor{ma}{RGB}{140,86,75}

\definecolor{in}{RGB}{148,103,189}

\definecolor{en}{RGB}{44,160,44}

\definecolor{di}{RGB}{255,127,14}

\definecolor{nd}{RGB}{227,119,194}

\definecolor{p1}{RGB}{56,113,193}
\definecolor{p2}{RGB}{254,109,23}
\definecolor{p3}{RGB}{92,186,86}
\definecolor{p4}{RGB}{220,68,87}

\newcommand{\latinphrase}[1]{\textit{#1}}  % always italic
\newcommand{\etal}{\latinphrase{et~al.}\xspace}

\definecolor{mulberry}{rgb}{0.77, 0.29, 0.55}

\usepackage[bookmarks,backref=true,linkcolor=black]{hyperref} %,colorlinks
\hypersetup{
	pdfauthor = {},
	pdftitle = {},
	pdfsubject = {},
	pdfkeywords = {},
	colorlinks=true,
	linkcolor= black,
	citecolor= black,
	pageanchor=true,
	urlcolor = black,
	plainpages = false,
	linktocpage
}

\ifCLASSINFOpdf
\else
\fi

\newcommand\designChoice{
	\begin{figure}
		\centering
		\includegraphics[width=\columnwidth]{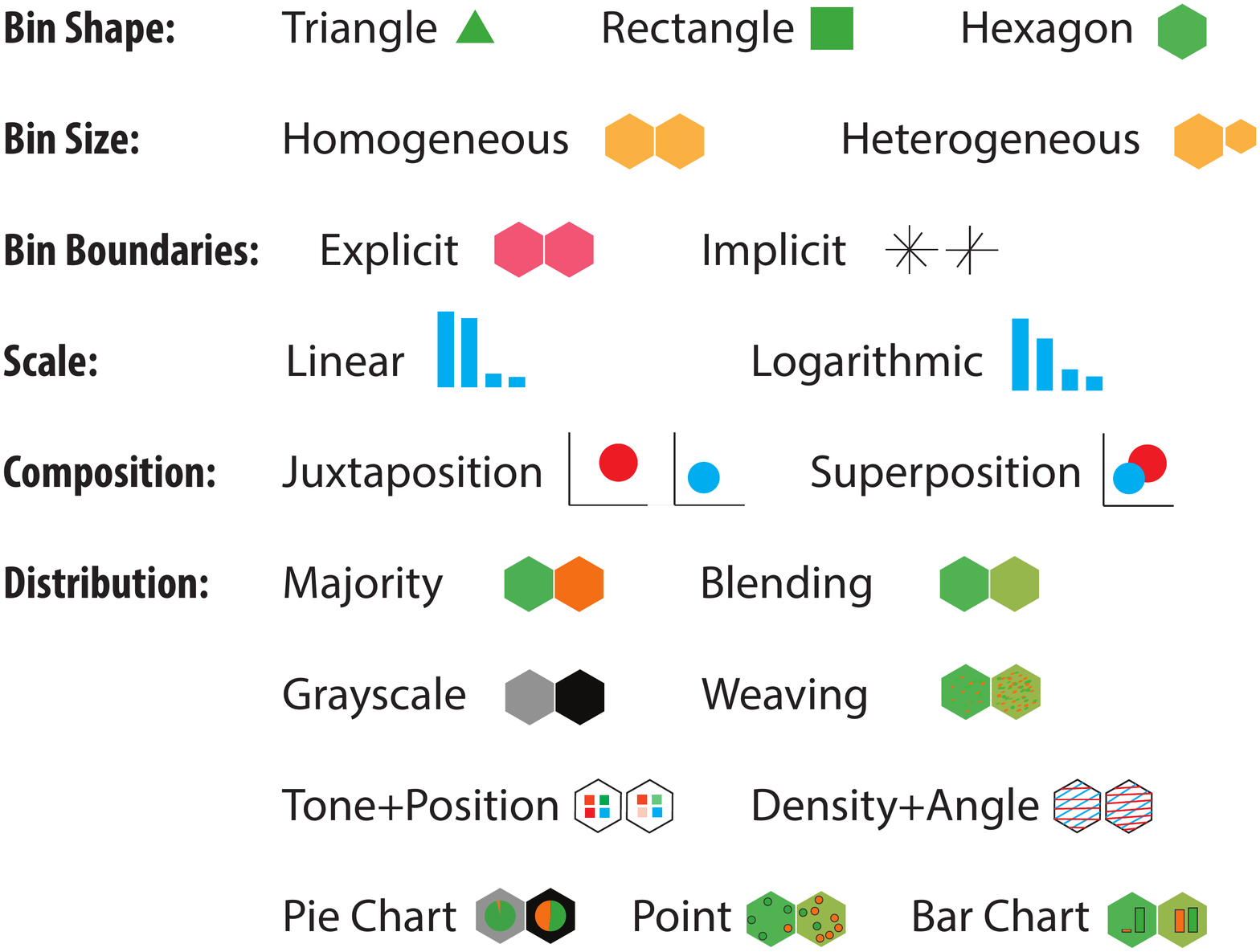}
		\caption{
			Design space for binned scatterplots.
			The first three categories are relevant for the representation of bins, while the last three categories deal with the representation of classes and class distributions.
		}
		\label{designChoice}
	\end{figure}
}

\newcommand\nbabinall{
	\begin{figure*}[t!]
		\centering
		\includegraphics[width=\textwidth]{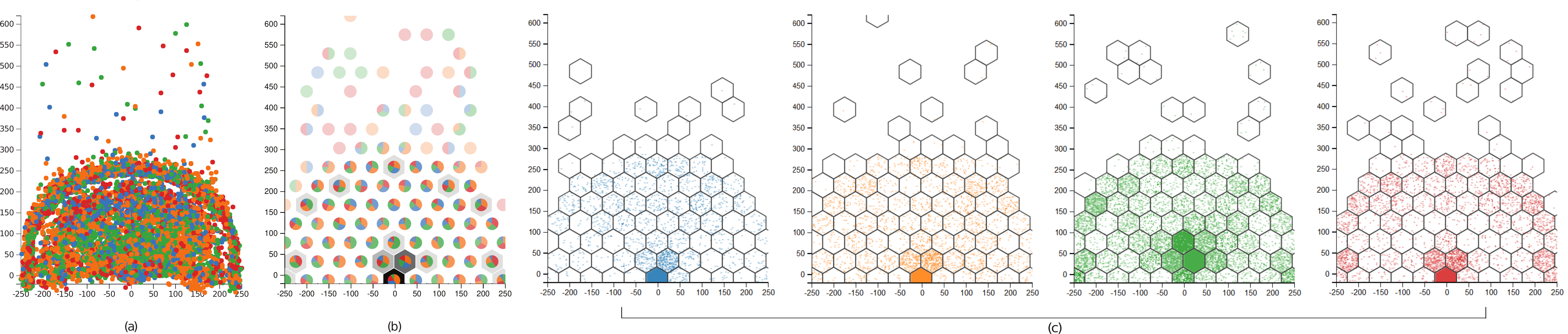}
		\caption{Shooting positions for four \emph{guards}, 1~\crule[p1]{0.25cm}{0.25cm},
			2~\crule[p2]{0.25cm}{0.25cm}, 
			3~\crule[p3]{0.25cm}{0.25cm}, and
			4~\crule[p4]{0.25cm}{0.25cm},
			from the NBA dataset.
			(a) only conveys rough densities.
			(b) is a design of binned aggregation that uses a grayscale color map in the background to show the total number of data points in each bin.
			In addition, each bin includes a pie chart that shows class proportions in that bin.
			(c) is a juxtaposition-based design of binned aggregation.
			Each class is shown in a separate plot.
			Binned aggregation lets users discern much more information from the data than the tradition scatterplot that has overplotting issues.}
		\label{nbabinall}
	\end{figure*}
}

\newcommand\treecoverall{
	\begin{figure*}[t!]
		\centering
		\includegraphics[width=\textwidth]{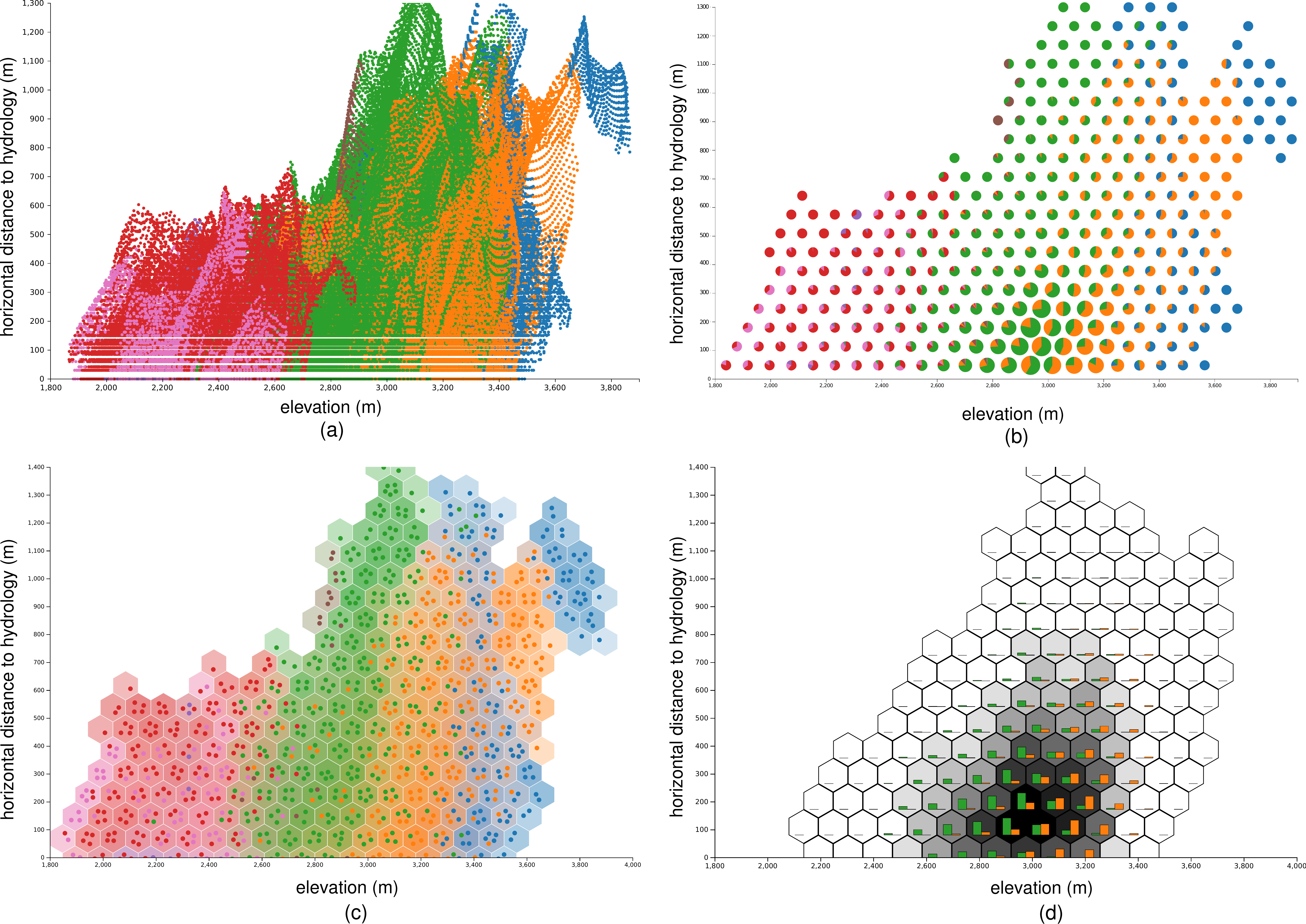}
		\caption{
			These plots are based on the Tree Cover in Colorado dataset.
			It has seven different tree types: \crule[spruce]{0.25cm}{0.25cm} spruce and fir,
			\crule[lodgepole]{0.25cm}{0.25cm} lodgepole pine,
			\crule[ponderosa]{0.25cm}{0.25cm} ponderosa pine,
			\crule[cottonwood]{0.25cm}{0.25cm} cottonwood / willow,
			\crule[aspen]{0.25cm}{0.25cm} aspen,
			\crule[douglas]{0.25cm}{0.25cm} douglas fir, and
			\crule[krumm]{0.25cm}{0.25cm} krummholz.
			In the scatterplot (a), details about the distribution of classes are hard to discern due to overplotting.
			The second visualization (b) is created using binned aggregation.
			It allows to compare bin density encoded by the pie sizes.
			In addition, class diversity is also shown by the pie charts.
			The third visualization (c) shows class identity within each bin and provides better information about minority classes.
			The fourth visualization (d) shows distributions for two classes (\crule[lodgepole]{0.25cm}{0.25cm} and \crule[spruce]{0.25cm}{0.25cm}) based on a bar chart design.
			It allows us to compare their respective distributions.
		}
		\label{treecoverall}
	\end{figure*}
}

\newcommand\ldaall{
	\begin{figure*}[t!]
		\centering
		\includegraphics[width=1.0\textwidth]{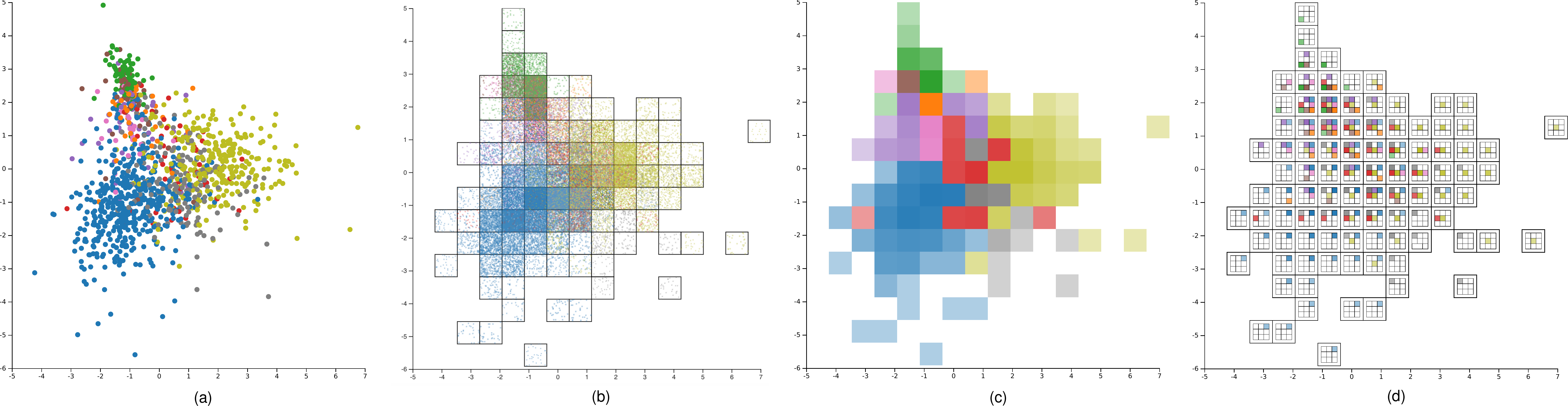}
		\caption{
			These plots are based on the Early Modern Drama corpus.
			It comprises nine types of dramatic texts:
			\crule[tr]{0.25cm}{0.25cm} tragedy,
			\crule[tc]{0.25cm}{0.25cm} tragicomedy,
			\crule[co]{0.25cm}{0.25cm} comedy,
			\crule[hi]{0.25cm}{0.25cm} history,
			\crule[ma]{0.25cm}{0.25cm} masque,
			\crule[in]{0.25cm}{0.25cm} interlude,
			\crule[en]{0.25cm}{0.25cm} entertainment,
			\crule[di]{0.25cm}{0.25cm} dialogue, and
			\crule[nd]{0.25cm}{0.25cm} non-dramatic.
			In the scatterplot (a), details about the distribution of classes and overlap are hard discern due to overplotting.
			The second plot (b) shows rough densities of the entire dataset based on color weaving.
			The third visualization (c) shows rectangular bins of the data, color based on high class-internal frequency values that convey a sense of regions with density peaks for each class.
			(d) is a small multiple design per bin that encodes frequencies for each class showing regions of overlap and extensions of classes.
		}
		\label{ldaall}
	\end{figure*}
}

\newcommand\blendWeaveAndHatch{
	\begin{figure*}
		\centering
		\includegraphics[width=\textwidth]{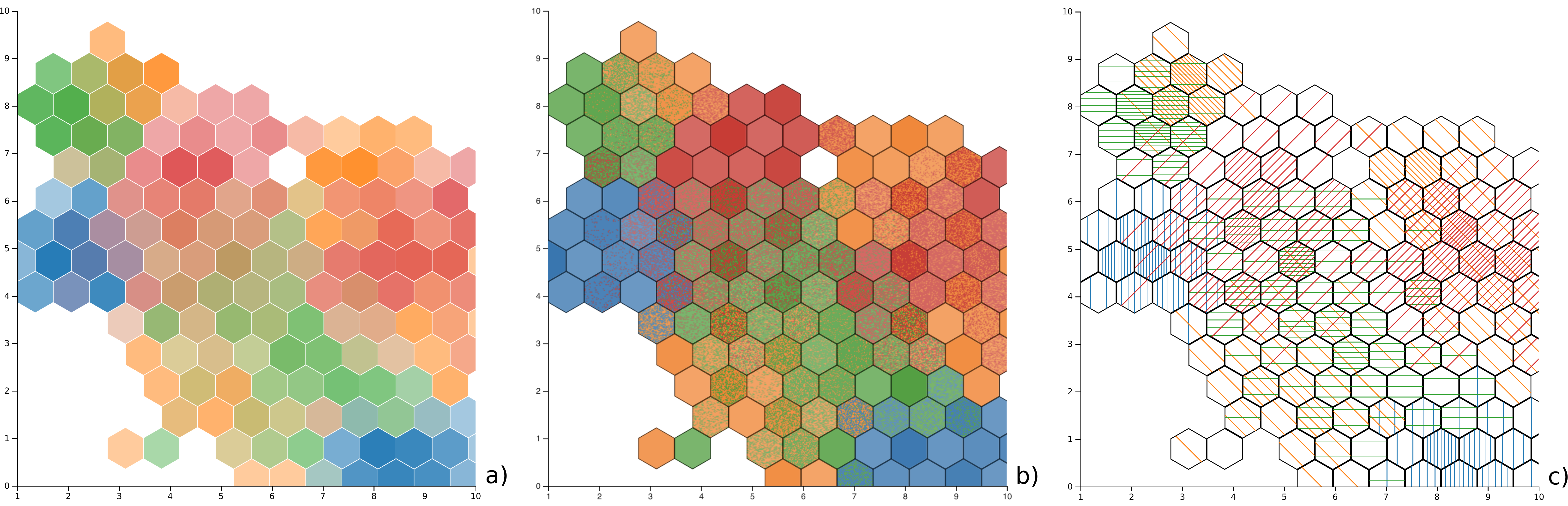}
		\caption{
			Three visualizations of an artificially generated dataset with four different classes (\classred, \classgreen, \classblue, \classorange).
			(a) demonstrates color blending, which uses binned color and saturation to encode class labels and their intensity.
			(b) demonstrates weaving with bin-internal normalization, resulting in fully filled bins with the number of colored fragments encoding bin-relative class intensities.
			(c) demonstrates hatching, which uses both angle and color to encode class labels.
		}
		\label{fig:blend_weave_and_hatch}
	\end{figure*}
}

\newcommand\designTaskGrid{
	\begin{table}[t!]
		\centering
		\includegraphics[width=\columnwidth]{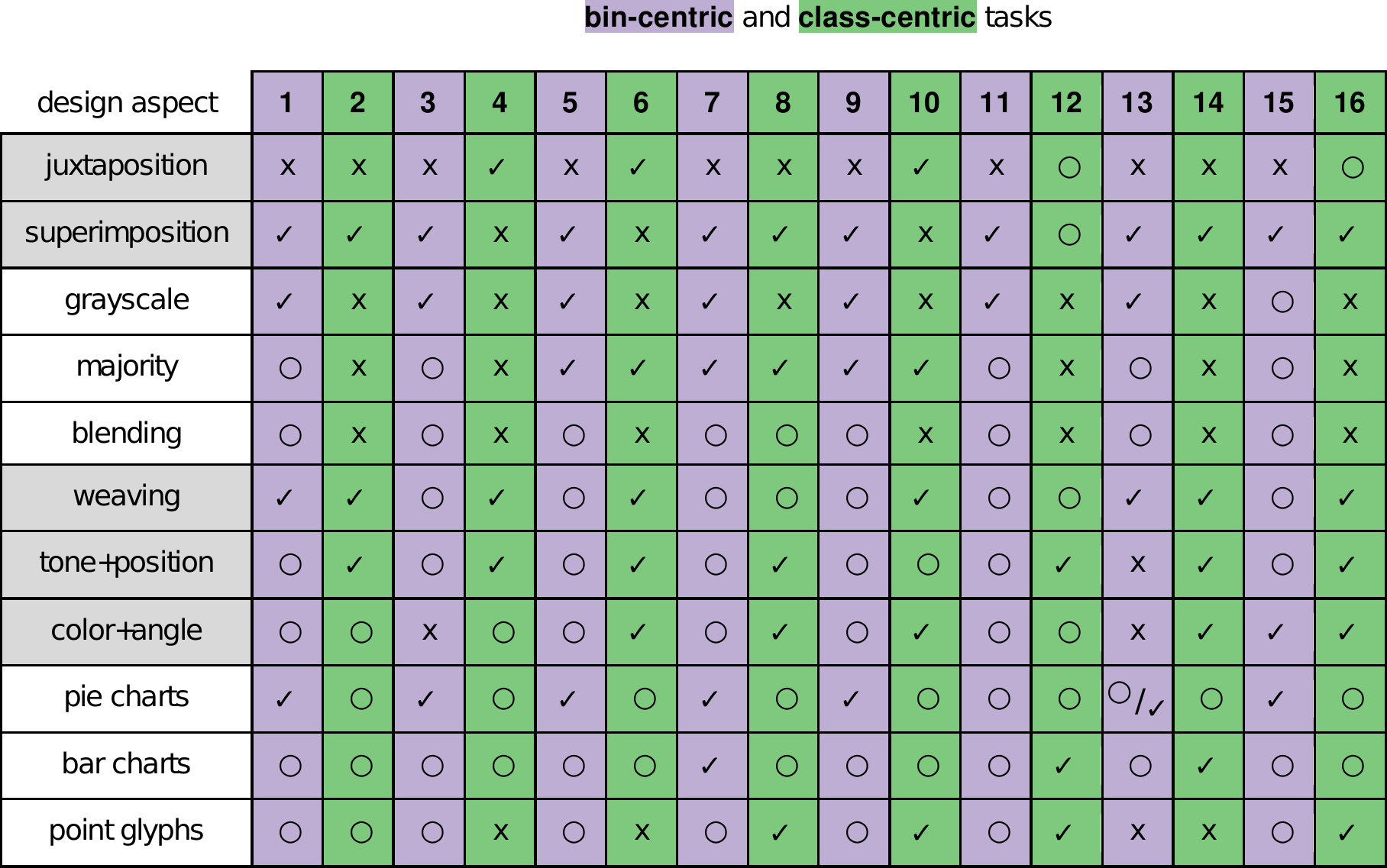}
		\caption{
			This table summarizes the discussion of Section~\ref{sec:design} by mapping design aspects to tasks.
			A check mark means a design is appropriate for the task; a circle means the design supports (aspects of) the task but may not be the best choice for it; an X means the design does not support the task.
		}
		\label{table:designTasks}
		\vspace{-6mm}
	\end{table}
}

\newcolumntype{K}{p{0.3cm}}
\newcolumntype{N}{>{\raggedright\arraybackslash}p{6cm}}
\newcommand\newTasksTable{
	\begin{table*}
		\begin{center}
			\begin{tabular}{@{}Klr@{\hskip4pt}Nr@{\hskip4pt}N@{}} \toprule
			 & Task	& \multicolumn{2}{l}{Bin-centric} & \multicolumn{2}{l}{Class-centric} \\ \midrule
			 & Explore neighborhood & 1 & Explore properties of bins in a neighborhood & 2 & Explore properties of classes in a neighborhood \\
			 & Search for known motif & 3 & Find known pattern across bins & 4 & Find known pattern across classes \\ 
			\multirow{-3}{*}{\adjustbox{margin=0 0 2.4mm 0,set height=0.3cm,angle=90}{browsing}} & Explore data & 5 & Unusual patterns within or across bins, global trends between bins & 6 & Unusual patterns within or across classes, global trends within or between classes \\ \midrule
			 & Characterize distribution & 7 & Do bins close to each other have similar properties? Or within a certain area or range of values? & 8 & Does a class occupy certain areas of the plot? Does its distribution have a particular shape? Do classes correlate in certain areas? \\
			 & Identify anomalies & 9 & Identify bins that are outliers based on the general distribution & 10 & Identify classes or subsets of classes that are outliers in a certain region \\
			 & Identify correlation & 11 & Determine level of correlation of bin properties along both dimensions & 12 & Determine level of correlation for class members along both dimensions \\
			 & Numerosity comparison & 13 & Compare density in different regions of the space & 14 & Compare class density in different regions of the space \\
			\multirow{-8}{*}{\adjustbox{margin=0 0 1mm 0,set height=0.3cm,angle=90}{aggregate-level}} & Understand distances & 15 & Understand a given spatialization and the coverage of the bins & 16 & Understand a given spatialization and the coverage of classes \\
			\bottomrule
			\end{tabular}
		\end{center}
		\caption{
			Tasks for binned scatterplots, based on the general scatterplot tasks compiled and categorized by Sarikaya and Gleicher~\cite{Sarikaya2018}.
			We reduced the original, larger set of tasks to those that capture high-level data characteristics according to the definition by Schulz~\etal~\cite{Schulz2013}.
			For binned scatterplots, each of the resulting tasks can have a \emph{bin-centric} or a \emph{class-centric} scope.
		}
		\label{targetTable}
		\vspace{-6mm}
	\end{table*}
}

\begin{document}

\title{Visual Designs for Binned Aggregation of Multi-Class Scatterplots}

\author{Florian Heimerl, Chih-Ching Chang, Alper Sarikaya, and Michael Gleicher,~\IEEEmembership{Member,~IEEE}
}

\IEEEtitleabstractindextext{
\begin{abstract}
Point sets in 2D with multiple classes are a common type of data.
A canonical visualization design for them are scatterplots, which do not scale to large collections of points.
For these larger data sets, binned aggregation (or binning) is often used to summarize the data, with many possible design alternatives for creating effective visual representations of these summaries.
There are a wide range of designs to show summaries of 2D multi-class point data, each capable of supporting different analysis tasks.
In this paper, we explore the space of visual designs for such data, and provide design guidelines for different analysis scenarios.
To support these guidelines, we compile a set of abstract tasks and ground them in concrete examples using multiple sample datasets.
We then assess designs, and survey a range of design decisions, considering their appropriateness to the tasks.
In addition, we provide a web-based implementation to experiment with design choices, supporting the validation of designs based on task needs.
\end{abstract}

}

\maketitle

\IEEEdisplaynontitleabstractindextext

\IEEEpeerreviewmaketitle

\IEEEraisesectionheading{\section{Introduction}\label{sec:intro}}

\IEEEPARstart{P}{oint} sets in a 2D space are a common type of data, with scatterplots as a canonical visualization solution.
The points that represent data items in these sets often have a categorical (or class) attribute.
For example, a geographic dataset might contain the coordinates of different types of places.
While regular scatterplots can accommodate multi-class data using a visual variable to encode class membership, such as color or shape~\cite{Alper2011, Collins2009, Dinkla2012, Meulmans2013}, they scale poorly to datasets with a large number of points~\cite{Keim2010}.
Overplotting can hinder analysis of the data due to point occlusion, inhibiting the identification of class attributes.

A range of approaches exist to address overplotting by using methods to abstract or summarize 2D point data (see \S\ref{sec:rel}).
Generally, we can describe abstracting a dense scatterplot as a two-step process.
During the first step, a density field is created to capture relevant distributional characteristics of the underlying dataset.
In a second step, the abstract density field is transformed into a visual representation that conveys those characteristics to users.

The method of choice for the first step is influenced by properties of the data and the analysis tasks that the designer seeks to support.
Designers can either choose continuous methods, such as kernel density estimation (KDE), or discrete methods.
The latter are called \emph{binning}, or \emph{binned aggregation}, because they split up the 2D space into a regular lattice of equally shaped and sized bins.
We do not necessarily advocate for binning relative to other aggregation methods.
However, since it is a popular method of choice (see \S\ref{sec:rel}), in this work, we focus on visual designs for it.
Binned aggregation results in a tessellation of the original scatterplot, i.e., an arrangement of bin shapes without gaps or overlaps.
Item frequencies are stored as a discrete scalar field defined over the regular lattice.
For multi-class data, the binning operation is either done once for each class, resulting in a separate bin structure per class, or once per data set, with separate statistics of class frequencies per bin.

Binning is used in a variety of analysis scenarios across different fields~\cite{Scott2015, Battersby2016}, with historic examples from the early 19\textsuperscript{th} century~\cite{Playfair1801}.
Its popularity seems to stem from being straightforward to interpret and easily adaptable during analysis, for example, to semantic properties of the data at hand.
Because of this, the selection of appropriate bin shape and size is often influenced by both, statistical criteria~\cite{Sturges1926} and perceptual effectiveness of the resulting visual encoding, as well as semantic data properties.

The second step of the abstraction process---translating the binned data to a visual representation---has a variety of designs published in the literature~\cite{Jo2019}.
In addition, design solutions for closely related problems, such as visualizing and comparing scalar fields over irregular tessellations (e.g., geopolitical entities in choropleth maps), can often be used directly or easily adapted to regular grids.
However, most of the current literature dealing with multi-class or multi-variate datasets focuses on a small set of very specific tasks, such as trend detection~\cite{Livingston2011b} or class composition in an area of interest~\cite{Bo2014}.
In addition, a recently published task and design space for scatterplots~\cite{Sarikaya2018} does not cover designs for binned aggregation at the necessary depth.
At the same time, visual analysis systems built for large data volumes~\cite{Li2014, Liu2013, Perrot2015}, including multi-class data, demonstrate the need for effective visualizations.

Due to its popularity, there is a clear need to explore and map the design space of binned aggregation visualizations for multi-class data.
We provide a mapping between design decisions and supported analysis tasks, which we derive from a collection of abstract tasks that users seek to perform when analyzing multi-class datasets.
Guided by existing designs and motivated by example analyses, we construct a list of design choices for binned aggregation designs.
We extend the choices collected from the literature with new designs that cover additional tasks.
Connecting the design choices back to the tasks, we then provide guidelines that can be used by visualization designers to create practical applications.
The overall contributions of this work are:
\begin{itemize}
	\item A compilation of representative analysis tasks for binned multi-class point data.
	\item A survey of existing visual designs for binned aggregation, applicable to multi-class analysis scenarios. We then extend these designs in a structured way.
	\item An assessment of design choices based on the tasks that can serve as design guidelines for visualization designers.
\end{itemize}

\section{Motivation and Example Data} \label{sec:motiv}
\label{sec:basketball}
\nbabinall
This section introduces an example dataset to motivate the use of binning and illustrate the influence of design aspects on task support.
Our list of tasks, that will be discussed in \S\ref{sec:tasks}, is listed in Table~\ref{table:designTasks}.
In the remainder of this section, we use forward references to the tasks in the task list.

The data comes from a publicly available National Basketball Association (NBA) dataset~\cite{nbadata2017}. 
We extract the 
shooting positions of four \emph{point-guard} players for a set of games.
Point-guards have the role of securing scoring opportunities for their teams, and thereby tend to shoot for the goal less, and pass to their teammates more than other players.
As an example, a potential analysis scenario may be that teams want to analyze shooting trends of opposing players to learn and exploit their opponents' strengths and weaknesses.

Figure~\ref{nbabinall} shows multiple representations of the data that each serve different tasks.
Figure~\ref{nbabinall}a is a traditional scatterplot.
Due to the high density close to the goal (bottom center), overdraw makes it difficult to learn about the distribution of shot positions.
We can still see that the area around the goal seems to be the densest area on the court.
Density then drops towards the 3-point line (semi-circle centered at the goal) and increases again close to the line, while the rest of the court seems to generally have a very low density.
If players hit the goal from behind the 3-point line, the team will get 3 instead of 2 points.

When analyzing attacking strategies of their opponents based on recorded shooting positions, a team might want to inspect the area around the goal, and the homogeneity of class distributions in that area (\emph{task 1: explore neighborhood -- bin}).
Figure~\ref{nbabinall}b shows a binned representation of the same data, using hexagonal bins that allows us to compare densities and per-bin class ratios across the space.
Overall density for each of the bins is encoded through their background color, using a grayscale color map with darker colors representing higher density.
Class distributions within the bins are shown as pie charts, with color indicating each player.
This provides insights into whether the probability that a certain player will shoot is similar for any position in this region, or whether it differs depending on the position relative to the goal.
From Figure~\ref{nbabinall}b, we can learn that there is high variation of point densities between the goal and the 3-point line, indicating that the players have different strategies in this region.

The next question is how exactly those patterns differ in this region by player (\emph{task 2: explore neighborhood -- class}), shifting from a bin-based to a class-based task.
For this, Figure~\ref{nbabinall}c shows another design alternative that juxtaposes density maps for each individual player.
It allows comparing between the players in detail.
The design is based on color weaving, and generates colored fragments in each bin according to the relative, class-based shot frequency of a player.
Compared to the previous design, this one focuses on density distributions within each class and allows to analyze and compare them across classes.
We can see that while three of the players have very similar patterns (\crule[p1]{0.25cm}{0.25cm}, \crule[p2]{0.25cm}{0.25cm}, and \crule[p4]{0.25cm}{0.25cm}), with the highest density in the bin surrounding the goal, \crule[p3]{0.25cm}{0.25cm} unusually tends to attack from the left side of the goal.

Next, the analyst of the team looks for clusters of specific patterns: bins across the space that have roughly equal class distributions (\emph{task 3: search motif -- bin }).
This marks regions on the court in which player strategies do not differ much.
Based on Figure~\ref{nbabinall}b, one such cluster is around the 3-point line and expands towards the goal at the center of the court.
Comparing this to the general density distribution across the court, with its peak around the goal and very low-density areas beyond the 3-point line (\emph{task 5: explore data -- bin}), we can see that there is a medium density region around the three point line.
One likely explanation for this is that players attempt shots from the 3-point line or the center of the court when under time pressure, resulting in this pattern of roughly equal distributions.
There are two additional regions with significant shot density: the two corners of the field.
We can see that \crule[p4]{0.25cm}{0.25cm} in particular, and \crule[p3]{0.25cm}{0.25cm} are the most likely players to shoot from there.
Switching back to a class-based perspective, we can see in Figure~\ref{nbabinall}c that both \crule[p1]{0.25cm}{0.25cm} and \crule[p2]{0.25cm}{0.25cm} are similar (\emph{task 4: search motif -- class}) in that most of their density accumulates around the goal, with the rest of the shots close to the 3-point line, and both players nearly never attempt shots from other regions.
On the other hand, \crule[p4]{0.25cm}{0.25cm} and \crule[p3]{0.25cm}{0.25cm} both have a significant number of shots from the left and right corner of the court.
In particular, \crule[p3]{0.25cm}{0.25cm} has the most unique pattern (\emph{task 6: explore data -- class}) with much more shots in the area between the 3-point line and the goal compared to the other three players.

\vspace{1mm}
\noindent \textbf{Additional Datasets}
and examples for all tasks are discussed in \S\ref{sec:tasks}.
We introduce datasets that differ in their interpretations of the dimensions, which influences the tasks that users might be interested in.
The data we have looked at in this section is spatial data, where the position of the points has a spatial analog (player position on a sports field).
In these types of datasets, absolute and relative positions across the space tend to be relevant for the analysis.
Other data sets have well-defined, continuous dimensions without an intrinsic spatial referent, such as the weight and gas usage of cars.
For this type of data, class and bin-centric distribution of densities and absolute position relative to the axes are relevant.
Thirdly, dimensions may have been created by dimensionality reduction, such as PCA or tSNE, and have no interpretable meaning.
Instead, relative positions, class distributions, and potential clusters in the resulting 2D space are relevant during analysis.

\section{Related Work} \label{sec:rel}
We discuss related work relevant to each of the two parts of the process of generating binned visual representations of scatterplots:
(1) the methods of binning and abstracting the data into a lattice, and
(2) the methods and techniques to visualize the lattice in a task-appropriate manner.
While we do not necessarily advocate for binning, we subsequently review some of the reasons why it remains popular.

\subsection{Data Aggregation Methods}
Binning is a frequently used method to help address the problem of overdraw with traditional scatterplots.
According to Battersby~\cite{Battersby2016}, its computational efficiency, and the perceptual advantages of regular grids for comparing densities within and between aggregated plots are reasons for the popularity of binning.
Cleveland describes how it can be used to address issues of overdraw and occlusion in his book~\cite{Cleveland1985}.  
He places great importance on the ability of the binned representation to convey a sense of aggregate distributions (see also discussion of choosing bin shape and size in \S\ref{sec:shape} and \S\ref{sec:size}), while sacrificing the ability of the viewer to identify individual points.  
Carr stresses the computational efficiency of binning, and reiterates its importance to show distributions by applying binning to SPLOMs in order to understand distributions across many dimensions~\cite{Carr1986}.

Aside from regular lattices, other shape types and tesselations can be used to achieve different design goals.  
Battersby \etal~\cite{Battersby2016} show that geographical distortion of bin shape can impact viewers' ability to compare count and density measures.  
An example method of methodologically creating irregular bins is to ensure that each bin contains the same number of data instances~\cite{Bak2009,Hao2010}. 
In this case, as the number of data items increases, spatial bins must also grow in size.  
As a follow-up, Hao \etal~\cite{Hao2010} propose variable binned scatterplots, a method that shows each region in a separate plot large enough to show all points without occlusion.  

Cartograms are an example of modifying bin size to communicate magnitude of data instances, and different techniques have varying analysis trade-offs~\cite{Sun2010, Nusrat2015}.  
For choropleth maps that communicate magnitude within bins with color, Brewer \etal~\cite{Brewer1997} evaluate the use of diverging, sequential, and spectral color schemes to support different analysis tasks.  

As an alternative to discrete aggregations methods, kernel density estimation (KDE) can aggregate point data to a continuous scalar field~\cite{Scott2015}.  
Techniques such as Splatterplots~\cite{Mayorga2013} can utilize these methods to create semantically meaningful visualizations that automatically scale to the available screen-space.  
While Splatterplots uses thresholding to denote areas of density, Chen \etal~\cite{Chen2014} takes advantage of space-aware subsampling to illustrate proportional densities with smaller amounts of points.  
Isocontours can also be computed from a KDE, describing densities much like a topographic map~\cite{Urness2003}. 
Jo \etal~\cite{Jo2019} describe how vonoroi tesselations can create variable bins shapes and sizes, dependent on the distribution.  

While our focus is on regular bins, we also consider tasks and designs for irregular lattices as sources for design ideas.

\newTasksTable

\subsection{Visual Representations for Binned Data}
The results of binning multi-class data are scalar fields over a regular lattice.
For the visual encoding of this type of data, many alternatives are available to communicate both frequencies and distributions of classes, and facilitate comparison between them based on both.
Color is one common visual variable to encode frequency.
Aside from using continuous color channels such as luminance to convey frequency~\cite{Brewer1997}, color ramps can be chosen to emphasize frequency peaks~\cite{Liu2013}, and quantized to improve viewers' accuracy of color perception and recall~\cite{Padilla2017}.
For particular tasks, designs may want to highlight frequencies that do not match an underlying distribution---Correll and Heer~\cite{Correll2017} uses Bayesian surprise as an orthogonal dimension in a bi-variate color map, emphasizing those areas with unusually high or low frequency.

Another common use of color is to communicate class identity.
Color weaving~\cite{Shenas2007} assigns proportions of color conveying membership pixel-wise to an area, then permutes those pixels to create a proportional tapestry of color.
It thus uses color to communicate both, class identity as well as frequency.
Color weaving can help to elucidate class proportions in dense areas of the plot, as described by Luboschik \etal~\cite{Luboschik2010}.
Attribute blocks~\cite{Miller2007} can show the proportions of a large number of classes by further subdividing bins, using each bin to encode a single class frequency via color.

Other choices for conveying frequency and class membership include texture and size, sometimes combined with a color encoding.
Ware~\cite{Ware2009} uses textons (small texture-based glyphs) and color to allow the viewer to compare two scalar fields.
Tobler~\cite{Tobler1973} uses textures that are able to convey continuous values in choropleth maps via quantization.

The ideas for visualizing regular and non-regular 2D density fields from these approaches helps to inform our exploration of visual designs suitable for binned aggregation throughout this paper.

\subsection{Linking Task and Design}
Taxonomies of tasks and designs use abstraction to emphasize differences and similarities without dependencies on the implementation and domain details.  
Tasks are a core consideration in the design of visualizations, scaffolding the viewer to obtain the desired information about the data (see Munzner~\cite{Munzner2015} for an overview).  
Early work by Shneiderman~\cite{Shneiderman1996} has been extended by numerous other task taxonomies; see Bremer and Munzner~\cite{Brehmer2013} for an overarching taxonomy.  
More domain-specific task sets have been proposed for areas such as cartography~\cite{Andrienko2006,Roth2013}. 
Of particular interest for this paper, Schulz \etal~\cite{Schulz2013} create a generalized space of visualization analysis task, taking into account the cardinality of objects being considered, as well as the types of data characteristics (e.g., distribution, outliers) communicated by the task.

Our interest is in how tasks relate to the designs of a resulting binned visualization.  
While no such work exists for this scenario, previous has focused on other combinations of design decisions and tasks.  
Javed and Elmqvist~\cite{Javed2012} describe the design space of compositing visualizations, basing their analysis on the literature.  
Schulz \etal~\cite{Schulz2011} review and extend the visualization design space for hierarchical data, similar as we do for multi-class binning.
Borgo \etal~\cite{Borgo2013} compile an overview of glyph-based visualization and design guidelines with associated tasks from examples in the literature.

Informed by the guidelines in Kerracher and Kennedy~\cite{Kerracher2017}, we validate our task classification through examining existing taxonomies and instantiating abstract tasks on concrete analyses.
Closely related to our paper, Sarikaya and Gleicher~\cite{Sarikaya2018} provide a space of analysis tasks, data characteristics, and design decisions derived from existing examples in the literature.
Their guidance is generalized to the entire space of scatterplot designs, suggesting a need for more specific analysis for particular scenarios such as for multi-class binning.
Jo \etal~\cite{Jo2019} generate a grammar for deriving numerous binned designs, highlighting decisions of encoding type and normalization.
This work, however, stops short of drawing relationships between design decisions and the types of analysis tasks they support.
We seek to fill this gap with this work.

\section{Tasks for Binned 2D Data} \label{sec:tasks}
In this section, we derive the task definitions listed in Table~\ref{targetTable} based on a set of abstract tasks.
We then ground each of these tasks in an example using a dataset.

\subsection{Task List}
\ldaall
Tasks for 2D point data that was aggregated by binning are connected to the more general tasks for regular scatterplots, for which an analysis and categorization of tasks has been published~\cite{Sarikaya2018}.
The task space that we derive and discuss in this section, is an extended subset of this larger collection of tasks that can be supported by general, unbinned scatterplots.
We refine the derived set of tasks according to the task design space of Schulz~\etal~\cite{Schulz2013}, and ground each of these abstract tasks in a concrete example from the previously introduced data sets.

Based on a review of relevant literature, Sarikaya and Gleicher~\cite{Sarikaya2018} collect and categorize a set of 12 tasks that users do with scatterplots.
Those tasks are grouped into three different categories, comprising \emph{object-centric} tasks, \emph{browsing}, and \emph{aggregate-level} tasks.
The first type, \emph{object-centric}, focuses on single data objects, and includes identifying and finding the location of a particular object.
In other words, \emph{object-centric} tasks cover all the low-level data characteristics of Schulz~\etal's task design space~\cite{Schulz2013}.
The second category, \emph{browsing}, comprises tasks focused on either single data items or higher level structures such as clusters, and thus targets low- as well as high-level data characteristics.
The third category, \emph{aggregate-level} tasks, focus entirely on high-level data characteristics.
When working with binned scatterplots, the data analyst has decided that aggregating the data is the best way to perform the task at hand.
Since the aggregation step abstracts away from single items, leaving only high-level data characteristics, we can reduce the set of potential tasks supported by a binned scatterplot to \emph{browsing} and \emph{aggregate-level} tasks.

Binned representations of multi-class data introduces two new visual elements that analysis tasks can target, \emph{bins} and \emph{classes}.
The dimension that captures this is called the \emph{scope} (or \emph{cardinality}) of a task~\cite{Schulz2013}.
Each of the tasks in our space can either be targeted at bins (\emph{bin-centric}), or at classes (\emph{class-centric}).
Extending the task set along this dimension is helpful for tasks supported by binned representation of 2D data, since it significantly influences the adequacy of designs to serve a task.
Table~\ref{targetTable} lists all resulting tasks and diversifies them into a \emph{bin-centric} and a \emph{class-centric} version.
In addition, a more extensive table, mapping all of the abstract tasks to high-level data characteristics and example tasks discussed in the following section is available as supplemental material (see also \S\ref{sec:dis} for a discussion of task completeness).

\subsection{Task Examples}\label{sec:example_tasks}
We have already seen examples for the first six tasks from Table~\ref{targetTable} back in \S\ref{sec:motiv}.
Here, we introduce two additional datasets and show designs and examples for the remaining tasks. 

\vspace{1mm} \noindent \textbf{Early Modern Drama Collection} contains the full text of 1,242 dramas from the years 1576--1700.
The texts are categorized into nine different genres, including \classtr tragedy, \classtc tragicomedy, and others (see caption of Figure~\ref{ldaall}).
Both dimensions have been generated using topic modeling to extract eight distinct topics based on the document-level co-occurrences of words across the corpus.
A topic is represented as a list of weighted words that are used in documents to talk about the topic.
We then picked two topics as dimensions to lay out the documents, based on the amount of words that each of the documents contains from the respective topic.
An example analysis scenario is to explore how those two topics separate the different drama genres.

Figure~\ref{ldaall}a shows a traditional scatterplot of the dataset based on those two dimensions.
While we are able to see some rough structure, especially in terms of class distribution for the larger classes such as for \crule[co]{0.25cm}{0.25cm}~comedies and \crule[tr]{0.25cm}{0.25cm}~tragedies, it is hard to spot the positions and distribution of the smaller classes.
To get a better sense of the general density distribution of the plot (\emph{task 7: characterize distribution -- bin}), we can take a look at the design in Figure~\ref{ldaall}b.
It uses color weaving to combine unnormalized color counts from all classes.
The number of colored fragments representing the overall density of each bin.
We can see that the general density is roughly distributed within a triangular shape, with density centers close to its three corners (\emph{task 9: identify anomalies -- bin}).

Next, we are interested in whether the different drama genres are separated based on the position within the plot (\emph{task 8: characterize distribution -- class}).
Figure~\ref{ldaall}c can give us initial insight into which bins are density centers for the classes.
It uses rectangular bins colored by interpolating the colors of all classes in each bin, weighted by densities normalized across categories.
Bins are thus dominated by the color of the class that has density peaks within the respective bin.
It thus provides an overview of class intensities across the space.
We can see that the two largest groups, \crule[co]{0.25cm}{0.25cm}~comedies and \crule[tr]{0.25cm}{0.25cm}~tragedies are mostly placed on the left and the right of the plot.

While Figure~\ref{ldaall}c colors entire bins based on color blending, Figure~\ref{ldaall}d shows density information for each class by mapping tone of each color to class intensity.
Rather than providing a coarse overview, it allows users to see more specific properties of class distributions and supports an in-depth analysis of class overlaps.
We can see that the density centers for the drama types \classtr tragedy and \classco comedy are situated along an axis from the left side of the space to its right side, with the density center for \classtc tragicomedy right between both.
All other types are roughly aligned along a perpendicular axis from the top of the plot further down.
We can also see, that while class density centers seem to be separated by the dimensions of this space, there is a huge overlap between the classes.
One interesting class whose distribution pattern is different from the others (\emph{task 10: identify anomalies -- class}) is \classhi history, which has its density center right in the middle of the plot and has lots of overlap with every other class in the dataset.

\treecoverall
\vspace{1mm} \noindent \textbf{Colorado Tree Coverage} dataset is part of the UCI machine learning data repository~\cite{Lichman2013}.
It contains data about the arbor environment of four wilderness areas in Colorado.
Each entry of the data set contains attributes of one individual tree, including variables such as tree type, position, and additional environmental factors.
Overall the data set covers seven different types of trees.
An example analysis scenario for this data set is to analyze how different environmental variables influence absolute and relative proportions of tree types in a region.
For this example, we focus on two specific variables from the data set.
The first one,`elevation', encodes the meters above sea level at which a tree grows, and helps to stratify the environment into different regions that tree types might prefer.
The second one, `horizontal distance to hydrology', encodes the number of meters (in horizontal direction) to the next water source.

Figure~\ref{treecoverall}a shows a traditional scatterplot of the roughly 600,000 trees.
We can get an impression of the areas in which different types of trees grow, but overdraw is a huge problem.
A question such as whether there is a correlation between elevation level and the variety of tree types that grow on that level (\emph{task 11: identify correlation -- bin}) is hard to answer.
From Figure~\ref{treecoverall}b, however, we can easily see that variety decreases with rising elevation based on the number of classes within the pie charts.
The design is based on hexagonal bins without showing explicit boundaries or bin shapes.
This design aspect has the advantage of reducing visual clutter, but it also makes it harder to read exact bin positions off the plot.
Pie charts convey class proportions in each of the bins, while the area of each pie chart encodes overall point density in the bin.
From this design, we can easily learn about the general density distribution across the bins in addition to a rough impression of relative class distributions.
Analyzing distributions within single classes, and the regions of overlap between multiple classes are, however, hard to explore with this design.

Figure~\ref{treecoverall}c is an alternative for this data set with explicit outlines for each bin.
The background colors are the result of blending together all class colors present in a bin, weighted according to their frequency within the bin.
This enables us to quickly read the main class present in a bin and gauge each bin's pureness with respect to class distribution.
Similar to the previous design, the plot conveys rough class distributions based on the blended background colors.
This design also shows a subset of points for each bin, sampled based on the distribution of classes within the bin, but at least one data point per class.
It allows us to see classes that have very low frequency compared to others.

From Figure~\ref{treecoverall}c we can see that the most prominent type of trees in an environment varies with elevation level.
We then wonder whether elevation level can help us separate between different tree types (\emph{task 12: identify correlation -- class}).
While there is a lot of overlap generally between types, we can see from Figure~\ref{ldaall}c, for example, that \classponderosa ponderosa pines do not grow above 2800m, while \classkrumm krummholz starts to appear at about 3300m.
Figure~\ref{ldaall}b also allows us to quickly spot regions of high (or low) density---such as the high concentration of trees at 2800--3200 meters, very close to the nearest water source (\emph{task 13: numerosity comparison -- bin}).

The next questions we might ask is whether both tree types in the area contribute equally to this peak, or whether this is due to one particular type (\emph{task 14: numerosity comparison -- class}).
From unnormalized frequencies in Figure~\ref{treecoverall}d, we can see that there are actually two overlapping density peaks in \classlodgepole lodgepole pines, and, at slightly higher altitude levels \classspruce spruce/fir that both combine to this peak in tree density.
Another question we might have for the dataset are general boundaries of tree growth (\emph{task 15: understand distances -- bin}).
From Figure~\ref{treecoverall}c we can see that \classcottonwood cottonwood, \classdouglas douglas firs, and and \classponderosa ponderosa pines need water close by (700 meters to the nearest water source), while the other four types are much less dependent on close surface water.
The plot also shows that most trees have a preferred elevation range in which they grow, with bands of roughly 500-600 meters of altitude and large overlaps between the different tree types across those regions (\emph{task 16: understand distances -- class}).

\section{Design Space} \label{sec:design}
To collect designs for binned scatterplots, we searched for existing solutions from publications in the visualization and cartography fields.
We then systematically created extensions and adaptions of them to cover additional data and tasks.
For each of the designs discussed that have been used previously, we refer to relevant publications.
In addition, we also discuss studies and other sources that provide insight about the effectiveness of design aspects.

\designChoice

\subsection{Representing Bins}
The first three design dimensions discuss choices to generate and represent bins and their properties within the 2D space.

\subsubsection{Shape} \label{sec:shape}
All of the designs that we discuss are based around visual representations of the bins generated during the first part of the abstraction process.
For this reason, the choice of bin shape, which is part of the output of the first step of binning, depends on statistical and distributional characteristics of the data, as well as perceptual properties of the resulting visualization.
In addition to choosing an effective shape for the data and task at hand, designers should also take care in choosing the scale of axes, as the perception of distribution can be affected by the positioning of bins.

There are only three shapes to tessellate a 2D space: triangles, rectangles, and hexagons.
Scott~\cite{Scott1988} suggests that triangular bins should be avoided since dividing the space up into triangles results in a higher expected per-point position error than alternatives.
In addition, triangles require rotations of the shape.
Rectangular bins provide good contrast between orthogonal and diagonal neighbors~\cite{Birch2006}, making them a good choice if the alignment with vertical or horizontal neighbors are important.
They are the only shape with a constant interval along both axes, making them a good choice if either the task requires reading off intervals with a certain precision from the plot, or the chosen intervals have semantic meaning (such as temporal or spatial units).
Hexagons are particularly common on maps, and are considered aesthetically superior to alternatives~\cite{Carr1992}.
They are better at representing local neighborhoods of bins~\cite{Birch2007}, making them a good choice for bin-centric tasks that focus on local structure, such as \emph{tasks 1 and 2: explore neighborhood -- bin/class}.
Hexagons also introduce the least expected error between a point and the bin center, resulting in the least expected distortion of density counts~\cite{Scott1988}.
They are thus also well-suited for tasks that involve the identification of fine-grained local density gradients, such as \emph{tasks 9 and 10: identify anomalies -- bin/class}, or \emph{tasks 11 and 12: identify correlation -- bin/class}.

\subsubsection{Size} \label{sec:size}
Similar to bin shape, its size is also influenced by data characteristics and visual properties of the resulting display.
Its choice determines the number of bins, limiting or enhancing the spatial fidelity of the visualization.
For designs that are based on multiple plots (see discussion about comparison in \S\ref{sec:composition}), bin size can either be homogeneous or heterogeneous across the plots.
The latter allows designers to choose different spatial resolutions for each class, but complicates the mapping between plots (more in \S\ref{sec:interaction}).

Binning creates a 2D histogram of the data space, where bin size controls the degree of aggregation that is applied to the data.
It determines what details users are able to discern about the data.
Methods that find optimal solutions for a large range of different datasets have been studied by Wand~\cite{Wand1997} and Knuth~\cite{Knuth2006}.

Another aspect of bin size are perceptual aspects of the visual representations of the bins, which is governed by the available screen space for the visualization.
Smaller bin sizes reduces the fidelity of communicating both spatial and class-proportional information within each bin---this greatly affects class-specific tasks, such as \emph{task 14: numerosity comparison -- class}, or bin-centric ones that compare inbalanced class proportions, such as \emph{bin-centric tasks 1: explore neighborhood, 5: explore data, or 7: characterize distribution}).
The available space also affects color perception~\cite{Stone2012}.
Therefore, there is a balance in trade-offs between maximizing the number of bins to reduce spatial aliasing (smaller bin size), but large enough to convey distributional information for each bin (larger bin size).

\subsubsection{Bin Boundaries} \label{sec:boundary}
Bins that have no explicit boundaries have to contain glyphs to communicate distributional information (much like QTonS~\cite{Ware2009}, an overlay for a scalar field), such as the pie glyphs in Figure~\ref{treecoverall}b. 
An advantage of boundary-less designs is reduced clutter.
For this reason, depending on the complexity and visual properties of the glyphs, smaller glyph sizes can be accommodated which are beneficial to tasks that profit from high spatial resolution (e.g., \emph{tasks 9 and 10: identify anomalies -- bin/class, and 11 and 12: identify correlation -- bin/class}).
However, missing bin boundaries make it harder to gauge the exact area a bin covers, hindering tasks that depend on this (e.g., \emph{tasks 7 and 8: characterize distribution -- bin/class}).

An example for a design that explicitly encodes the spatial boundaries of bins is shown in Figure~\ref{ldaall}b.
In addition to making bin intervals easier to read off the plot, explicit boundaries also help with mapping bins across different plots~\cite{Tory2003}, for example when using juxtaposed designs (as discussed in \S\ref{sec:composition}).
Another advantage of bin boundaries is that they introduce bins as separate visual elements into the plot, making it easier to support additional user interaction with them (see \S\ref{sec:interaction}).

\subsection{Composition}\label{sec:composition}
In this section, we discuss encoding classes and class distributions for each of the bins, and ways of composing this class-specific information into a multi-class density map.

\vspace{1mm}
\noindent \textbf{Class Identity}
The designs we discuss encode bin position in the data space as position in the plot.
Thus, while position would be a very salient channel for encoding identity~\cite{Mackinlay1986}, it is already in use for the two primary data attributes.
From the remaining choices, Livingston~\etal~\cite{Livingston2012} find that color is quite effective.
This is by far the most popular choice in the literature, with few historic exceptions that use texture to encode class identity~\cite{Pierce1894}.
All of our designs and the following discussions are based on using color to encode class identity.
Those colors are combined in different ways to communicate a range of class distribution properties.
There are two exceptions to this, discussed in \S\ref{sec:colorplus}: one of our designs (attribute blocks) uses relative position within the bins, while another (hatching) uses angle to encode class identity.
Both designs redundantly encode these additional visual variables with color for class identity.

\vspace{1mm}
\noindent \textbf{Normalization}
After creating the spatial bins, the data items in each of the bins are reduced to raw frequency counts for each class present in the bin.
Normalization of those raw counts and the visual encoding of the resulting distribution are independent.
Still, choice of normalization has a significant influence on the adequacy of visual encodings and the types of comparisons that those encodings need to support for the task at hand.
The details of these dependencies are discussed with the designs in \S\ref{sec:distribution}.
There are three options for normalization:
\begin{itemize}
	\item \emph{bin-internal}: all frequency counts are normalized by the maximum frequency in their bin.
	\item \emph{class-internal}: all frequency counts are normalized by the maximum frequency in their class.
	\item \emph{global}: all frequency values are normalized by the overall maximum frequency across bins and classes.
\end{itemize}

Designs based on bin-internal normalizations favor bin-centric tasks (\emph{odd-numbered tasks}).
As an example, Figure~\ref{treecoverall}b has been created using bin-internal normalization, with each of the pie charts depicting bin-relative class distributions.
It lets us compare overall densities and their distribution between bins and clusters of bins.
Comparing class intensities and numerosity of a particular class across the space, however, is not possible with this design.
Conversely, class-centric tasks (\emph{even-numbered tasks}) are supported with class-internal normalization that allows comparing class intensities across bins.
In addition, some class-centric tasks require global normalization, such as numerosity-based \emph{task 14: numerosity comparison -- class}, which depends on the ability to compare raw frequencies across bins.
Figure~\ref{ldaall}b and c are both use class-internal normalization and allow to compare class-specific properties.
Figure~\ref{treecoverall}d supports a class-centric task with global normalization, and allows for both comparison across and within bins.

\vspace{1mm}
\noindent \textbf{Scale}
Many of the designs are not effective at displaying details at the tail end of distributions.
For example, with unnormalized frequencies, if a large number of data items fall into smaller bins, it is difficult to see any details outside those dense bins.
The same is generally true for class distributions within bins.
Figure~\ref{nbabinall}b uses grayscale color to encode the number of total points in each bin, with darker background indicating higher numerosity.
Since the density of data items is particularly high around the basket, relative color difference of other bins across the space are diluted due to their relatively smaller difference in numerosity.
One possible solution is to scale the raw count numbers for each bin based an attenuation function.
The log-function is a popular choice~\cite{Cleveland1985b}.
Different variants are discussed in the literature as dynamic range reduction, see Shirley and Marschner~\cite{Shirley2009} for examples.

\vspace{1mm}
\noindent \textbf{Comparison}
The tasks from Table~\ref{targetTable} can be considered comparison tasks.
The bin-centric version of each task is a visual comparison between bin properties, while the class-centric versions are comparisons between local or global class properties.
Choice of visual design to support these comparisons has a significant effect on the types of tasks supported by a visualization.
Two comparison designs~\cite{Gleicher2011} widely used for binned scatterplots are juxtaposition and superimposition.
The former shows multiple visualization components separately, one per class in the data set (such as Figure~\ref{nbabinall}c).
Consequently, each class has its own coordinate system that can be adapted (and the resulting space aggregated) independently of the others. 
Superimposition overlays multiple visualizations, one for each class, into a single view.
In this case, all data points share a common coordinate system.
Distributional information about classes is mixed on the bin level for superimposition-based designs (Figure~\ref{ldaall}c is an example for this).

Juxtaposition-based designs can only be combined with global or class-based normalization.
This makes them ideal for the analysis of density distributions within classes, and comparison of global features of those distributions across classes, such as \emph{class-centric tasks 4: search motif, 6: explore data , and 10: identify anomalies}.
One advantage of juxtaposed designs is that overall less information has to be presented in a bin, making it easier to find a good balance between proportional and spatial fidelity while remaining readable.
This also accommodates smaller bin sizes, resulting in potentially higher spatial resolution of the visualizations.

However, it is difficult to link classes across bins between multiple juxtaposed plots.
While this is true for plots with the same scale and bins, plots that differ in those properties make it even harder to visually match regions across multiple plots.
For this reason, tasks that require users to compare local features of density distributions across classes, such as \emph{class-centric tasks 2: explore neighborhood, 8: characterize distribution, and 14: numerosity comparison}, are better served with a superimposition-based design.
Juxtaposed designs neither support any of the bin-centric tasks well, because all of them require to collect densities across classes for each bin.
This is in line with Livingston~\etal~\cite{Livingston2012}, who find that when tasks require reading multiple variables across different plots, juxtaposition has a higher error rate and slower response time.

\subsection{Density and Class Distribution}\label{sec:distribution}
This section discusses designs that convey classes, class proportions, and distributions within bins.

\subsubsection{Single Color}\label{sec:single_color}
These design alternatives are methods for using bin background color to communicate properties of the class distribution.

\vspace{1mm}
\noindent \textbf{Luminance (Grayscale or Color)}
While a univariate encoding does not communicate class distribution, it can convey item density within a bin.
Color luminance imparts an implicit order of magnitude~\cite{Munzner2015}.
This allows users to locate and explore bin properties related to item densities, and thus can serve a number of bin-centric tasks (e.g., \emph{tasks 1: explore neighborhood, 3: search motif, 5: explore data, 7: characterize distribution, 9: identify anomalies, 11: identify correlation, and 13: numerosity comparison}).
Padilla~\etal~\cite{Padilla2017} find that binned color scales, as opposed to continuous ones, expedite tasks that include the identification of maxima (such as \emph{bin-centric tasks 5: explore data, 9: identify anomalies, and 13: numerosity comparison}).
Adding glyphs to the foreground of bins can extend task coverage of the base design (as in Figure~\ref{nbabinall}b, for example).
For class-based tasks, luminance can also be used in a juxtaposition-based design to convey densities of single classes separately.
To encode class identity, class colors are best used in each of the juxtaposed multiples instead of grayscale colors.

\vspace{1mm}
\noindent \textbf{Color of Majority Class}
Another option is to only show the color of the majority class for each bin.
Considering the choice of normalization, there are two versions of this, with bin-internal (or global normalization), or class-internal normalization.
With the former, each bin is colored according to the class that is most prominent (in terms of item numerosity) within a bin.
This is useful for \emph{bin-centric tasks 5: explore data, 7: characterize distribution, and 9: identify anomalies} that target the general distribution and anomalies.
The latter version colors bins based on which class has the highest relative proportion of members in a bin.
Figure~\ref{ldaall}b shows an example of this, revealing class-specific density centers.
Generally, the design serves tasks that target distribution and potential anomalies within classes, such as \emph{class-centric tasks 6: explore data, 8: characterize distribution, and 10: identify anomalies}.
The example in Figure~\ref{ldaall}b uses an additional luminance encoding of overall item densities for each bin.
This basic design is particularly effective in scenarios with a small set of classes.

\vspace{1mm}
\noindent \textbf{Color Blending}
With color blending (Figure~\ref{fig:blend_weave_and_hatch}a), class colors are combined according to a weighted average.
Again, weights can be based on class-internal or bin-internal normalization to focus on bin-centric and class-centric tasks, respectively.
Interpretation of blended colors is generally hard in cases with more than two base colors~\cite{Gama2014}.
Color blending is thus most useful for data sets with a small number of classes, and little overlap between more than two of them.
Compared to the previous method, it can provide a sense of the bin purity in terms of the classes it contains, helping to support some bin-centric (e.g., \emph{task 1: explore neighborhood}) and class-centric tasks (e.g., \emph{task 8: characterize distribution}).

\subsubsection{Color + Additional Variables}\label{sec:colorplus}
This section discusses color-based designs that use additional visual variables.
These differ from the previous three designs by occupying the entire bin area to convey distributional information, and can thus not be complemented with additional visualizations.

\vspace{1mm}
\noindent \textbf{Color Weaving}
An alternative to color blending is color weaving, a technique that permutes the positions of colored fragments while maintaining proportionality of each color.
It takes advantage of the visual system’s ability to summarize color within an area~\cite{Albers2014}.
Hagh-Shenas~\etal~\cite{Shenas2007} show that users are better at discerning different contributing colors when using color weaving compared to color blending.
As a consequence, color weaving is as a more effective and scalable option to show class diversity and summary information for bins.
This, however, comes at the cost of not being able to combine this design with glyphs to cover additional tasks, because they would cover large areas of the space that are needed for accurate interpretation of the weaving pattern.

Depending on the normalization, color weaving can also be effective at conveying overall bin densities (as demonstrated in Figure~\ref{ldaall}b).
Bin-internal normalization will result in the optimal use of a bin's area to encode its class distributions by filling all available fragments (but does not allow comparisons across bins).
Global normalization, on the other hand, which allows for comparison of class frequencies across bins, results in some of the fragments of each bin remaining white.
Overall bin densities are then encoded by the ratio of white to colored fragments, allowing users to effectively compare overall densities across bins.
Figure~\ref{fig:blend_weave_and_hatch}b shows an example.
In addition to a variety of class-centric tasks, this design thus also supports many of the bin-centric tasks discussed for the luminance-based design in \S\ref{sec:single_color}.
Finally, we found that the third, class-internal, normalization option is not a good choice in combination with color weaving and may confuse users due to the implicit part-whole metaphor of color weaving.
This aspect of weaving is similar to pie chart glyphs, discussed in \S\ref{sec:glyphs}.
However, weaving can also be used in juxtaposed designs to generate multiple single-class density maps for class-centric tasks, such as in Figure~\ref{nbabinall}c.
One general disadvantage of weaving is that it is not very effective at conveying particularly small class proportions, since a low number of equally colored fragments is hard to discern.

\blendWeaveAndHatch

\vspace{1mm}
\noindent \textbf{Attribute Blocks: Tone and Position}
Attribute blocks~\cite{Miller2007} assign each class a square bin within a glyph.
Color luminance encoding a summary value for the class (see Figure~\ref{ldaall}d).
Livingston~\etal~\cite{Livingston2012} show that attribute blocks have a high error rate for tasks that require reading multiple variates.
Their experiments only use position for class membership, with a color gradient for magnitude.
In our experience, using additional class colors (similar to Miller~\cite{Miller2007}) makes extracting class-specific distributions easier.
While attribute blocks can be combined with any type of normalization, they excel at class-centric tasks (using class-internal normalization), in particular those that involve analyzing class overlap across the space (\emph{class-centric tasks 6: explore data, 8: characterize distribution, 12: identify correlation, and 16: understand distances}).

\vspace{1mm}
\noindent \textbf{Hatching: Color and Angle}
Hatching is a particularly common choice to encode numerosity in monochrome choropleth maps~\cite{Peterson1979}.
A viable option for multi class data is to use simple, angled strokes for different classes, varying their density to encode class intensity.
Livingston~\etal~\cite{Livingston2012} show that using orientation of lines to encode variable values (representing classes by angle) is a perceptually efficient, easily separable encoding for multiple variables.
In addition, it shows 
An example of multi class hatching is shown in Figure~\ref{fig:blend_weave_and_hatch}c, with class-internal normalization.
It encodes class identity as both color and angle.

Despite the general trade-off between line thickness and resolution of density values, the design keeps classes separable due to redundant encoding, even if colors are hard to identify for thin lines~\cite{Stone2014}.
A disadvantage is that the drawing order of the lines influences their prominence across bins---denser lines exacerbate this effect.
One solution is adding interaction that lets users change class drawing order (as discussed in \S\ref{sec:interaction}).
This design is most effective for low to medium density values, and thus works best with global or class-internal normalization, especially for data that is well distributed throughout the space.
In addition, comparisons within classes and across bins are supported due to redundant encoding of class identity, making the design well-suited for class-centric tasks that analyze and compare class distributions (\emph{class-centric tasks 6: explore data, 8: characterize distribution, 10: identify anomalies}).

\subsubsection{Glyphs}\label{sec:glyphs}
Glyphs are small, independent visual objects that encode attributes of a data record~\cite{Borgo2013}.
They can either be added to the foreground of bins, or used in isolation.
When added to a base design, they increase task coverage in bins whose background is based on a single color value, while their effectiveness is reduced with more complex backgrounds by placing high cognitive load on users~\cite{Ware2013}.
We discuss three alternative glyph designs.
While they cover all of the tasks to convey class distributions, this list could be extended by additional glyphs that encode arbitrary properties of the data instances in a bin, thus, e.g., covering potential additional domain specific tasks of a particular dataset.

\vspace{1mm}
\noindent \textbf{Part-Whole (Pie Charts)}
Pie charts are a widely used solution to show proportion of different classes and part-whole relationships.
They are a common visualization technique available in geographic information systems to show distributions on top of choropleth maps~\cite{Anselin1999}.
There are multiple variations of pie charts~\cite{Kosara2010}, including donut and square pie charts.
Both regular pie and donut charts have comparable perceptual effectiveness~\cite{Skau2016}.
Figure~\ref{nbabinall}b shows standard pie charts, with color ordered according to class proportion and combined with a grayscale density map.
An alternative to encode overall item numerosity per bin is to use pie area, as shown in Figure~\ref{treecoverall}b.
While this has the advantage of reducing the number of visual elements in the plot, it exacerbates the poor visibility of sparse classes in the pies.
Pie charts can only be used with bin-internal normalization due to their part-whole metaphor, and there is evidence that they outperform alternatives when comparing class proportions across bins~\cite{Spence1991}.
This makes them a good solution for \emph{bin-centric tasks 3: search motif, 5: explore data, 7: characterize distribution, and 9: identify anomalies}.
However, care should be taken in case a tasks involves the identification of clusters across the space~\cite{Lewandowsky1993}.

\vspace{1mm}
\noindent \textbf{Baseline Proportional (Bar Charts)}
An alternative glyph, also borrowed from choropleth maps~\cite{Few2009}, are small bar charts, as shown in Figure~\ref{treecoverall}d.
They come in two flavors: regular and stacked.
Bar charts can be effective at conveying relative class proportion within each bin, and have the advantage of having a common bar baseline.
They have been used to successfully visualize multivariate datasets~\cite{Bo2014}, and allow users to accurately read values of proportion when the bin size is large enough.
Similar to pie charts, it can be hard to perceive class colors when the bars are small, though this problem can be somewhat alleviated by ordering bars according to class labels.
For stacked bar charts, users can have difficulties determining precise class proportions~\cite{Kosara2010}.
Despite these disadvantages, bar charts are a versatile design choice that can be used with all three types of normalization.
In contrast to pie charts, they allow for comparison across bins, which makes them suitable for class-centric tasks.
Bar charts are particularly useful for comparisons across bins and classes (e.g., for \emph{task 14: numerosity comparison -- class}).

\vspace{1mm}
\noindent \textbf{Points}
Point glyphs emphasize class variance within the bins.
They maintain a balance between showing class proportions and the spatial distribution of different classes.
For this, we sample the points in a bin to retain features of the underlying distribution.
This sampling strategy can be combined with all three normalization methods.
Bertini~\etal~\cite{Bertini2006} and Chen~\etal~\cite{Chen2014} show that sub-sampling points helps to overcome the problem of overdraw while preserving much of the spatial information.
Figure~\ref{treecoverall}c shows an example of this.
We sample points based on the class proportion and reduce overlap as much as possible.
In the case that a sampled point overlaps with the boundaries of the bin, its position is moved slightly towards the center.
To represent all of the classes present in a bin, proportion of a class may be distorted during sampling since we show at least one sample for each existing class.
This helps to identify small frequency phenomena, particularly for class-centric tasks, such as \emph{tasks 12: identify correlation, and 16: understand distances}.

\subsection{Interaction}\label{sec:interaction}
In addition to adding glyphs, another way of extending the supported tasks of a given design is adding interaction features to it.
Designs can support interaction on classes, bins, or both.
Selection of classes can either be done via the legend, by letting users hover over and click on labels.
Alternatively, in case a design includes glyphs, which have visually distinct elements for each class in a bin, interaction with these elements can highlight or select the respective class.
Selection of a class can either show additional information, such as distribution of the class or overlaps with other classes, filter all visual elements of a certain class, e.g., showing only the bars for one class in bar chart glyphs.
For designs that depend on the order of classes (Hatching in \S\ref{sec:colorplus}), highlighted or selected ones can be moved to the foreground.
Alternatively, it could open another view of the data that focuses on the selected class (similar to a juxtaposition-based design \S\ref{sec:composition}), switching to a class-centric visualization.
Generally, supporting filtering by class extends the tasks supported by a given design with additional, class-centric tasks.

Similar to classes, interacting with bins can also show additional information about each bin (e.g., raw counts of class items in a bin).
In addition, for juxtaposition-based designs, interaction can help users identify corresponding areas across multiple plots.
Corresponding referents can all be highlighted if multiple plots are based on identical bins.
If this is not the case, an additional overlay on the other plots can be used that shows the spatial extent of the selected bin across the other plots, conveying class and item overlaps with the existing bins of each juxtaposed plot.
Another mode of interaction on the bin-level is zooming.
Based on the zoom-level, once the items that can be viewed on the available screen space no longer overlap each other, the summary design could switch to a detail view that shows item per bins.
This can provide additional support for class- and bin-centric tasks that depend on local details, such as \emph{tasks 1: explore neighborhood -- bin, 2: explore neighborhood -- class, and 8: characterize distribution -- class}.

\section{Implementation} \label{sec:impl}
We provide a web-based implementation to show interactive examples of the entire space of designs.
This allows designers to try out different design combinations, and make decisions based on comparing alternatives for a particular dataset.
Our framework is based on vue.js and vuetify for UI and interaction elements.
The visualizations are implemented using D3.js~\cite{Bostock2011}.
At the core, the implementation is built around a vue.js component that is made available as an open source, reusable software element\footnote{to be published with camera ready version of this submission} to include binned scatterplots in web visualization systems.

In addition, we provide a running web demo\footnote{http://graphics.cs.wisc.edu/Vis/binning/} of our implementation, that can be used to quickly try out and compare designs for new data sets.
Users can upload their own data, and interactively adjust design decisions and variables to adapt and modify designs and compare results.
The demo covers most of the design space discussed in \S\ref{sec:design}.
It is entirely browser-based, and scales to datasets consisting of many thousands of items, utilizing svg-based rendering in the browser.

\section{Discussion and Outlook} \label{sec:dis}
\designTaskGrid
Table~\ref{table:designTasks} summarizes the previous sections, mapping designs with respect to relevant tasks (\S\ref{sec:tasks}).
Design-task combinations with a checkmark work well to support the respective task.
We assume the best possible configuration of the design for the tasks; for example, correct normalization to optimally support the tasks.
The ``O'' mark identifies that a task is generally supported by a design, but either not all aspects of it are equally well supported, or there are better solutions for the task.
Cells with an ``X'' are combinations where the task is unsupported.
There is an exception: the \emph{pie charts} cell for \emph{task 13: numerosity comparison -- bin} has two symbols, for the version without, and the version with variable area to encode overall density.

The table groups designs into multiple categories, marked by alternating white or gray backgrounds.
While the first category (\emph{juxtaposition} vs. \emph{superposition}) can be combined with either the second or third category, the fourth one that contains glyphs for bins only work for superimposed designs.
In addition, two aspects from the two groups with a white background can be combined.
The upper one comprises all designs for bin background based on colors, while the lower one contains all additional glyphs to show distributions.
For example, point glyps support a variety of class-based tasks, and could be a good addition to an otherwise bin-centric background design for the bins if additional class-centric task are to be covered.

The summary table does not include the design aspects for representing bins, discussed in \S\ref{sec:design}.
This includes bin shape and size, which are dependent on a complex interplay of data characteristics and tasks.
For those, there is no generally straightforward mapping from designs to tasks, and designers should carefully evaluate and test potential visual variables against their data.
Resulting designs should clearly communicate relevant patterns in the data, while still assuring visual scalability and minimize potential aliasing effects.
Another design dimension excluded from the table is scale, whose connection to tasks is similarly complex.
Finally, the type of normalization is also not included, because it derives directly from the type of task, and is comprehensively discussed in \S\ref{sec:composition}.

In \S\ref{sec:tasks} we have derived a task space for binned representations of scatterplots, and grounded them in concrete examples.
An extended table of these tasks, containing a description of the concrete tasks is available as supplemental material.
In addition, the extended table maps each of the abstract tasks to one or more of the \emph{high-level task characteristics} defined by Schulz~\etal~\cite{Schulz2013}.
The table shows how our task space covers all of those characteristic.
Our set of task is thus comprehensive enough to cover all possible concrete tasks users may want to perform with a scatterplot using binned aggregation.
Furthermore, basing our tasks on an existing taxonomy that we adapted and extended, assures validity of the resulting task list~\cite{Kerracher2017}.
One category listed among the characteristics is the identification of \emph{outliers}.
As discussed previously, we have both bins and classes as visual objects present in our designs.
Identifying outliers thus targets either bins that differ on a local or global level from their surroundings, or classes that differ from other classes, depending on whether the task at hand is class- or bin-centric.

Our discussion only focuses on regular, fixed-sized latices.
The reason for this is that we are aiming at uncovering statistical and distributional properties of the data, which regular lattices help to preserve.
Using an adaptive grid method to create heterogeneous bins within one plot is likely to distort these properties.
Two popular methods often used to tessellate 2D data spaces are Voronoi tessellation and quad trees.
Rather than uncovering and visualizing distributional properties of underlying data, both structures adapt to data distributions with the goal of making certain computational operations on the data more efficient.
While creating designs to visually analyze and compare these data structures might be interesting, we consider it out of the scope of this work.

Another relevant aspect of the visualizations for binned aggregation is scalability, whether by computational or visual means.
We do not address aspects of computational efficiency in this work.
For highly scalable implementations, designers must assure that aggregation of data and combination of class-specific distributions are handled effectively.
Recently, Jo~\etal~\cite{Jo2019} described a general data model and processing pipeline that allows highly scalable implementations of the designs discussed in this work.
They use a 3-step processing model that uses an initial binning step to abstract and compress data, making subsequent processing more efficient.
Cottam et al.~\cite{Cottam2014} use binning in a similar fashion to increase the performance of interactive visualization on high-density data.

Whether a given design is visually scalable enough to analyze a given dataset has two main facets to it.
One is scalability with the number of data items, which is also dependent on the task and the distribution of the data.
Suitable designs should assure that relevant phenomena (for example correlations as in \emph{tasks 11 and 12: identify correlation -- bin/class}) are visually identifiable in the aggregated representation of the data.
This includes choosing correct spatial resolution and corresponding visual bin properties (such as bin boundaries) to minimize clutter and aliasing effects in the resulting visualization.
Depending on the available screen size and resolution, choosing appropriate designs for effective analysis might not be possible.
In this case, additional interaction, including zooming and panning within the data space are suitable methods to increase scalability.

The second facet is visual scalability with number of classes.
Based on the discussion of encoding class identity in \S\ref{sec:composition}, color is the most effective visual variable available to encode class identity.
However, this limits class scalability to roughly 10 classes, limiting the effectiveness of visual representations for datasets with a larger number of classes.
Some of the designs discussed are most effective with an even lower number of classes.
One way around this limitation is adding additional interaction to select subsets of classes to analyze.
Another possibility to increase class scalability is by adding additional glyphs that do not encode class identity, but rather features of the distribution of classes within each bin.
However, no good solutions for this problem exists.
We thus consider scalable visual designs for both regular and aggregated scatterplots that scale to a medium to high number of data classes an opportunity for future research.

We map alternative designs to tasks and rate their suitability.
Table~\ref{table:designTasks} shows that there is still significant overlap between designs and supported tasks (there are multiple check marks in many columns).
While the table still helps designers make good choices based on the overall set a design covers, for many of those alternatives, it remains unclear if there are actual gradual differences in the effectiveness for a tasks.
This could include general perceptual effectiveness, but also single aspects of a task that might be more efficiently solved with one of the design alternatives.
Given these overlaps, another interesting avenue for future research is the evaluation of sets of these designs for a set of tasks through user studies.
For this, our task and design analysis serve as a systematic foundation to select interesting task and design combinations.	

\section{Conclusion} \label{sec:conclusion}
We have explored the visualization design space for binned aggregation, a popular method to aggregate large multi-class scatterplots.
The main challenge for creating effective binned aggregation designs is the large space of design aspects.
Successful solutions that combine these aspect depend on tasks and data characteristics.
After discussing example datasets to motivate the problem, we derive a task space for abstract scatterplot visualizations and ground those abstracts tasks in concrete examples.
We then discuss design decisions for aggregated scatterplot visualizations, and their suitability to the tasks, and finally summarize these discussions by providing concrete guidelines for designers.
In addition, we also release an open source vue.js component that implements most design aspects discussed in this work to let designers quickly implement their own version of binned aggregation for scatterplots.

\ifCLASSOPTIONcaptionsoff
  \newpage
\fi

\bibliographystyle{abbrv}
\bibliography{bibliography}

\begin{thebibliography}{10}

\bibitem{Albers2014}
D.~Albers, M.~Correll, and M.~Gleicher.
\newblock Task-driven evaluation of aggregation in time series visualization.
\newblock In {\em Proceedings of the 2014 ACM annual conference on Human
  Factors in Computing Systems}, pages 551--560. ACM, May 2014.

\bibitem{Alper2011}
B.~Alper, N.~Riche, G.~Ramos, and M.~Czerwinski.
\newblock Design study of linesets, a novel set visualization technique.
\newblock {\em IEEE Transactions on Visualization and Computer Graphics},
  17(12):2259--2267, Dec 2011.

\bibitem{Andrienko2006}
N.~Andrienko and G.~Andrienko.
\newblock {\em Exploratory analysis of spatial and temporal data: a systematic
  approach}.
\newblock Springer Science \& Business Media, 2006.

\bibitem{Anselin1999}
L.~Anselin.
\newblock Interactive techniques and exploratory spatial data analysis.
\newblock {\em Geographical Information Systems: principles, techniques,
  management and applications}, 1:251--264, 1999.

\bibitem{Bak2009}
P.~Bak, M.~Schaefer, A.~Stoffel, D.~A. Keim, and I.~Omer.
\newblock Density equalizing distortion of large geographic point sets.
\newblock {\em Cartography and Geographic Information Science},
  36(3):237–250, 2009.

\bibitem{Battersby2016}
S.~E. Battersby, D.~Strebe, and M.~P. Finn.
\newblock Shapes on a plane: evaluating the impact of projection distortion on
  spatial binning.
\newblock {\em Cartography and Geographic Information Science}, 2016.

\bibitem{Bertini2006}
E.~Bertini and G.~Santucci.
\newblock Give chance a chance: modeling density to enhance scatter plot
  quality through random data sampling.
\newblock {\em Information Visualization}, 5(2):95--110, 2006.

\bibitem{Birch2006}
C.~P. Birch.
\newblock Diagonal and orthogonal neighbours in grid-based simulations:
  Buffon's stick after 200 years.
\newblock {\em Ecological Modelling}, 192(3--4):637--644, 2006.

\bibitem{Birch2007}
C.~P. Birch, S.~P. Oom, and J.~A. Beecham.
\newblock Rectangular and hexagonal grids used for observation, experiment and
  simulation in ecology.
\newblock {\em Ecological Modelling}, 206(3--4):347--359, 2007.

\bibitem{Bo2014}
S.~Bo.
\newblock Multivariate spatial visualization using geoicons and image charts.
\newblock Master's thesis, The University of Western Ontario, 2014.

\bibitem{Borgo2013}
R.~Borgo, J.~Kehrer, D.~H. Chung, E.~Maguire, R.~S. Laramee, H.~Hauser,
  M.~Ward, and M.~Chen.
\newblock Glyph-based visualization: Foundations, design guidelines, techniques
  and applications.
\newblock In {\em Eurographics (STARs)}, pages 39--63, 2013.

\bibitem{Bostock2011}
M.~Bostock, V.~Ogievetsky, and J.~Heer.
\newblock D3 data-driven documents.
\newblock {\em IEEE Transactions on Visualization and Computer Graphics},
  17(12):2301--2309, Dec. 2011.

\bibitem{Brehmer2013}
M.~Brehmer and T.~Munzner.
\newblock {A multi-level typology of abstract visualization tasks}.
\newblock {\em IEEE transactions on visualization and computer graphics},
  19(12):2376--85, dec 2013.

\bibitem{Brewer1997}
C.~A. Brewer, A.~M. MacEachren, L.~W. Pickle, and D.~Herrmann.
\newblock Mapping mortality: Evaluating color schemes for choropleth maps.
\newblock {\em Annals of the Association of American Geographers},
  87(3):411--438, 1997.

\bibitem{Carr1986}
D.~B. Carr, R.~J. Littlefield, and W.~L. Nichloson.
\newblock Scatterplot matrix techniques for large n.
\newblock In {\em Proceedings of the Seventeenth Symposium on the Interface of
  Computer Sciences and Statistics on Computer Science and Statistics}, pages
  297--306, New York, NY, USA, 1986. Elsevier North-Holland, Inc.

\bibitem{Carr1992}
D.~B. Carr, A.~R. Olsen, and D.~White.
\newblock Hexagon mosaic maps for display of univariate and bivariate
  geographical data.
\newblock {\em Cartography and Geographic Information Systems}, 19(4):228--236,
  1992.

\bibitem{Chen2014}
H.~Chen, W.~Chen, H.~Mei, Z.~Liu, K.~Zhou, W.~Chen, W.~Gu, and K.~L. Ma.
\newblock Visual abstraction and exploration of multi-class scatterplots.
\newblock {\em IEEE Transactions on Visualization and Computer Graphics},
  20(12):1683--1692, Dec 2014.

\bibitem{Cleveland1985b}
W.~S. Cleveland.
\newblock {\em The Elements of Graphing Data}.
\newblock Wadsworth Publ. Co., Belmont, CA, USA, 1985.

\bibitem{Cleveland1985}
W.~S. Cleveland, R.~McGill, et~al.
\newblock Graphical perception and graphical methods for analyzing scientific
  data.
\newblock {\em Science}, 229(4716):828--833, 1985.

\bibitem{Collins2009}
C.~Collins, G.~Penn, and S.~Carpendale.
\newblock Bubble sets: Revealing set relations with isocontours over existing
  visualizations.
\newblock {\em IEEE Trans. on Visualization and Computer Graphics (Proc. of the
  IEEE Conf. on Information Visualization)}, 15(6):1009,1016, 2009.

\bibitem{Correll2017}
M.~Correll and J.~Heer.
\newblock Surprise! bayesian weighting for de-biasing thematic maps.
\newblock {\em IEEE transactions on visualization and computer graphics},
  23(1):651--660, 2017.

\bibitem{Dinkla2012}
K.~Dinkla, M.~J. van Kreveld, B.~Speckmann, and M.~A. Westenberg.
\newblock Kelp diagrams: Point set membership visualization.
\newblock {\em Computer Graphics Forum}, 31(3pt1):875--884, 2012.

\bibitem{Few2009}
S.~Few and P.~Edge.
\newblock Introduction to geographical data visualization.
\newblock {\em Visual Business Intelligence Newsletter}, pages 1--11, 2009.

\bibitem{Gama2014}
S.~Gama and D.~Gonçalves.
\newblock Studying color blending perception for data visualization.
\newblock In N.~Elmqvist, M.~Hlawitschka, and J.~Kennedy, editors, {\em EuroVis
  - Short Papers}. The Eurographics Association, 2014.

\bibitem{Gleicher2011}
M.~Gleicher, D.~Albers, R.~Walker, I.~Jusufi, C.~D. Hansen, and J.~C. Roberts.
\newblock {Visual comparison for information visualization}.
\newblock {\em Information Visualization}, 10(4):289--309, 2011.

\bibitem{Shenas2007}
H.~Hagh-Shenas, S.~Kim, V.~Interrante, and C.~Healey.
\newblock Weaving versus blending: a quantitative assessment of the information
  carrying capacities of two alternative methods for conveying multivariate
  data with color.
\newblock {\em IEEE Transactions on Visualization and Computer Graphics},
  13(6):1270--1277, Nov 2007.

\bibitem{Hao2010}
M.~C. Hao, U.~Dayal, R.~K. Sharma, D.~A. Keim, and H.~Janetzko.
\newblock Variable binned scatter plots.
\newblock {\em Information Visualization}, 9(3):194--203, 2010.

\bibitem{Javed2012}
W.~Javed and N.~Elmqvist.
\newblock Exploring the design space of composite visualization.
\newblock In {\em 2012 IEEE Pacific Visualization Symposium}, pages 1--8, Feb
  2012.

\bibitem{Jo2019}
J.~Jo, F.~Vernier, P.~Dragicevic, and J.-D. Fekete.
\newblock A declarative rendering model for multiclass density maps.
\newblock {\em IEEE Transactions on Visualization and Computer Graphics},
  25(1), 2019.
\newblock To appear.

\bibitem{Cottam2014}
P.~W. Joseph A.~Cottam, Andrew~Lumsdaine.
\newblock Abstract rendering: out-of-core rendering for information
  visualization.
\newblock In {\em Proc. SPIE}, volume 9017, pages 9017 -- 9017 -- 13, 2014.

\bibitem{Keim2010}
D.~A. Keim, M.~C. Hao, U.~Dayal, H.~Janetzko, and P.~Bak.
\newblock Generalized scatter plots.
\newblock {\em Information Visualization}, 9(4):301--311, Dec. 2010.

\bibitem{Kerracher2017}
N.~Kerracher and J.~Kennedy.
\newblock Constructing and evaluating visualisation task classifications:
  Process and considerations.
\newblock In {\em Computer Graphics Forum}, volume~36, pages 47--59. Wiley
  Online Library, 2017.

\bibitem{Knuth2006}
K.~H. {Knuth}.
\newblock {Optimal Data-Based Binning for Histograms}.
\newblock {\em ArXiv Physics e-prints}, May 2006.

\bibitem{Kosara2010}
R.~Kosara and C.~Ziemkiewicz.
\newblock Do mechanical turks dream of square pie charts?
\newblock In {\em Proceedings of the 3rd BELIV'10 Workshop: BEyond Time and
  Errors: Novel evaLuation Methods for Information Visualization}, BELIV '10,
  pages 63--70, New York, NY, USA, 2010. ACM.

\bibitem{Lewandowsky1993}
S.~Lewandowsky, D.~J. Herrmann, J.~T. Behrens, S.-C. Li, L.~Pickle, and J.~B.
  Jobe.
\newblock Perception of clusters in statistical maps.
\newblock {\em Applied Cognitive Psychology}, 7(6):533--551, 1993.

\bibitem{Li2014}
C.~Li, G.~Baciu, and Y.~Han.
\newblock Interactive visualization of high density streaming points with
  heat-map.
\newblock In {\em 2014 International Conference on Smart Computing}, pages
  145--149, Nov 2014.

\bibitem{Lichman2013}
M.~Lichman.
\newblock {UCI} machine learning repository, 2013.

\bibitem{Liu2013}
Z.~Liu, B.~Jiang, and J.~Heer.
\newblock immens: Real-time visual querying of big data.
\newblock In {\em Proceedings of the 15th Eurographics Conference on
  Visualization}, EuroVis '13, pages 421--430, Chichester, UK, 2013. The
  Eurographs Association John Wiley \& Sons, Ltd.

\bibitem{Livingston2011b}
M.~A. Livingston, J.~Decker, and Z.~Ai.
\newblock An evaluation of methods for encoding multiple 2d spatial data.
\newblock In {\em Proceedings of SPIE}, volume 7868, pages 78680C--78680C--12,
  2011.

\bibitem{Livingston2012}
M.~A. Livingston, J.~W. Decker, and Z.~Ai.
\newblock Evaluation of multivariate visualization on a multivariate task.
\newblock {\em IEEE Transactions on Visualization and Computer Graphics},
  18(12):2114--2121, Dec 2012.

\bibitem{Luboschik2010}
M.~Luboschik, A.~Radloff, and H.~Schumann.
\newblock A new weaving technique for handling overlapping regions.
\newblock In {\em Proceedings of the International Conference on Advanced
  Visual Interfaces}, AVI '10, pages 25--32, New York, NY, USA, 2010. ACM.

\bibitem{Mackinlay1986}
J.~Mackinlay.
\newblock Automating the design of graphical presentations of relational
  information.
\newblock {\em ACM Trans. Graph.}, 5(2):110--141, Apr. 1986.

\bibitem{Mayorga2013}
A.~Mayorga and M.~Gleicher.
\newblock Splatterplots: Overcoming overdraw in scatter plots.
\newblock {\em IEEE Transactions on Visualization and Computer Graphics},
  19(9):1526--1538, Sept 2013.

\bibitem{Meulmans2013}
W.~Meulemans, N.~H. Riche, B.~Speckmann, B.~Alper, and T.~Dwyer.
\newblock Kelpfusion: A hybrid set visualization technique.
\newblock {\em IEEE Transactions on Visualization and Computer Graphics},
  19(11):1846--1858, Nov 2013.

\bibitem{Miller2007}
J.~R. Miller.
\newblock Attribute blocks: Visualizing multiple continuously defined
  attributes.
\newblock {\em IEEE Computer Graphics and Applications}, 27(3):57--69, May
  2007.

\bibitem{Munzner2015}
T.~Munzner.
\newblock {\em Visualization Analysis \& Design}.
\newblock CRC Press, 2015.

\bibitem{nbadata2017}
N.~B.~A. (NBA).
\newblock Nba shooting position dataset, 2017.

\bibitem{Nusrat2015}
S.~Nusrat, M.~J. Alam, and S.~G. Kobourov.
\newblock Evaluating cartogram effectiveness.
\newblock {\em CoRR}, abs/1504.02218, 2015.

\bibitem{Padilla2017}
L.~Padilla, P.~S. Quinan, M.~Meyer, and S.~H. Creem-Regehr.
\newblock {Evaluating the Impact of Binning 2D Scalar Fields}.
\newblock {\em IEEE Transactions on Visualization and Computer Graphics},
  23(1):431--440, jan 2017.

\bibitem{Perrot2015}
A.~Perrot, R.~Bourqui, N.~Hanusse, F.~Lalanne, and D.~Auber.
\newblock Large interactive visualization of density functions on big data
  infrastructure.
\newblock In {\em 2015 IEEE 5th Symposium on Large Data Analysis and
  Visualization (LDAV)}, pages 99--106, Oct 2015.

\bibitem{Peterson1979}
M.~P. Peterson.
\newblock An evaluation of unclassed crossed-line choropleth mapping.
\newblock {\em The American Cartographer}, 6(1):21--37, 1979.

\bibitem{Pierce1894}
F.~E. Pierce.
\newblock The tenement-house committee maps, 1894.

\bibitem{Playfair1801}
W.~Playfair.
\newblock {\em The commercial and political atlas: representing, by means of
  stained copper-plate charts, the progress of the commerce, revenues,
  expenditure and debts of england during the whole of the eighteenth century}.
\newblock T. Burton, 1801.

\bibitem{Roth2013}
R.~E. Roth.
\newblock An empirically-derived taxonomy of interaction primitives for
  interactive cartography and geovisualization.
\newblock {\em IEEE Transactions on Visualization and Computer Graphics},
  19(12):2356--2365, Dec 2013.

\bibitem{Sarikaya2018}
A.~Sarikaya and M.~Gleicher.
\newblock Scatterplots: Tasks, data, and designs.
\newblock {\em IEEE Transactions on Visualization and Computer Graphics},
  24(1):402--412, Jan. 2018.

\bibitem{Schulz2011}
H.~J. Schulz, S.~Hadlak, and H.~Schumann.
\newblock The design space of implicit hierarchy visualization: A survey.
\newblock {\em IEEE Transactions on Visualization and Computer Graphics},
  17(4):393--411, April 2011.

\bibitem{Schulz2013}
H.~J. Schulz, T.~Nocke, M.~Heitzler, and H.~Schumann.
\newblock A design space of visualization tasks.
\newblock {\em IEEE Transactions on Visualization and Computer Graphics},
  19(12):2366--2375, Dec 2013.

\bibitem{Scott1988}
D.~W. Scott.
\newblock A note on choice of bivariate histogram bin shape.
\newblock {\em Journal of Official Statistics}, 4:47--51, 1988.

\bibitem{Scott2015}
D.~W. Scott.
\newblock {\em Multivariate density estimation: theory, practice, and
  visualization}.
\newblock John Wiley \& Sons, Inc, 2nd edition edition, 2015.

\bibitem{Shirley2009}
P.~Shirley and S.~Marschner.
\newblock {\em Fundamentals of Computer Graphics}.
\newblock A. K. Peters, Ltd., Natick, MA, USA, 3rd edition, 2009.

\bibitem{Shneiderman1996}
B.~Shneiderman.
\newblock The eyes have it: a task by data type taxonomy for information
  visualizations.
\newblock In {\em Proceedings 1996 IEEE Symposium on Visual Languages}, pages
  336--343, Sep 1996.

\bibitem{Skau2016}
D.~Skau and R.~Kosara.
\newblock Arcs, angles, or areas: Individual data encodings in pie and donut
  charts.
\newblock {\em Computer Graphics Forum}, 35(3):121--130, 2016.

\bibitem{Spence1991}
I.~Spence and S.~Lewandowsky.
\newblock Displaying proportions and percentages.
\newblock {\em Applied Cognitive Psychology}, 5(1):61--77, 1991.

\bibitem{Stone2012}
M.~Stone.
\newblock In color perception, size matters.
\newblock {\em IEEE Computer Graphics and Applications}, 32(2):8--13, March
  2012.

\bibitem{Stone2014}
M.~Stone, D.~A. Szafir, and V.~Setlur.
\newblock An engineering model for color difference as a function of size.
\newblock In {\em Color and Imaging Conference}, volume 2014, pages 253--258.
  Society for Imaging Science and Technology, 2014.

\bibitem{Sturges1926}
H.~A. Sturges.
\newblock The choice of a class interval.
\newblock {\em Journal of the American Statistical Association},
  21(153):65--66, 1926.

\bibitem{Sun2010}
H.~Sun and Z.~Li.
\newblock Effectiveness of cartogram for the representation of spatial data.
\newblock {\em The Cartographics Journal}, 47(1):12--21, 2010.

\bibitem{Tobler1973}
W.~R. Tobler.
\newblock Choropleth maps without class intervals?
\newblock {\em Geographical Analysis}, 5(3):262--265, 1973.

\bibitem{Tory2003}
M.~Tory.
\newblock Mental registration of 2d and 3d visualizations (an empirical study).
\newblock In {\em Proceedings of the 14th IEEE Visualization 2003 (VIS'03)},
  VIS '03, pages 49--, Washington, DC, USA, 2003. IEEE Computer Society.

\bibitem{Urness2003}
T.~Urness, V.~Interrante, I.~Marusic, E.~Longmire, and B.~Ganapathisubramani.
\newblock Effectively visualizing multi-valued flow data using color and
  texture.
\newblock In {\em Proceedings of the 14th IEEE Visualization 2003 (VIS'03)},
  VIS '03, pages 16--, Washington, DC, USA, 2003. IEEE Computer Society.

\bibitem{Wand1997}
M.~P. Wand.
\newblock Data-based choice of histogram bin width.
\newblock {\em The American Statistician}, 51(1):59--64, 1997.

\bibitem{Ware2009}
C.~Ware.
\newblock Quantitative texton sequences for legible bivariate maps.
\newblock {\em IEEE Transactions on Visualization and Computer Graphics},
  15(6):1523--1530, Nov 2009.

\bibitem{Ware2013}
C.~Ware.
\newblock {\em Information visualization: perception for design}.
\newblock Elsevier, 2013.

\end{thebibliography}

\begin{IEEEbiography}
	[{\includegraphics[width=1in,height=1.25in,clip,keepaspectratio]{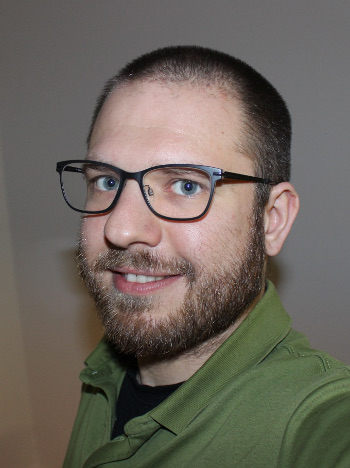}}]{Florian Heimerl}
	is a postdoctoral researcher in the visualization group at the University of Wisconsin--Madison. He received his PhD in computer science and his Diplom degree in computational linguistics from the University of Stuttgart, Germany. In Stuttgart, he was a doctoral student at the Institute for Visualization and Interactive Systems. His research interests include information visualization, visual analytics, and visual text analysis.
\end{IEEEbiography}

\begin{IEEEbiography}
	[{\includegraphics[width=1in,height=1.25in,clip,keepaspectratio]{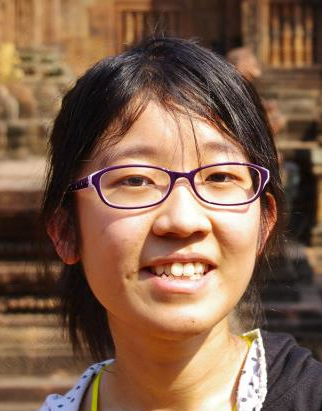}}]{Chih-Ching Chang}
	is a Software Engineer at Appier, Taiwan.  
  She received her Master degree in Computer Sciences from the University of Wisconsin--Madison. 
  Prior to joining the University of Wisconsin, Chih-Ching received her B.S. in 
  Electrical Engineering and Computer Sciences from the National Chiao Tung University.
\end{IEEEbiography}

\begin{IEEEbiography}
	[{\includegraphics[width=1in,height=1.25in,clip,keepaspectratio]{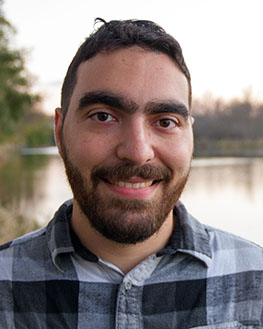}}]{Alper Sarikaya}
	is a Research and Development Engineer at Microsoft Corporation, working on Power BI.  He received his PhD in Computer Sciences from the University of Wisconsin---Madison.  His research focuses on how the design of summary visualizations can help focus relevant analyses and exploration for both specific and general visualization audiences.  Prior to joining the University of Wisconsin, Alper received his B.S. in Computer Science and Chemistry from the University of Washington.
\end{IEEEbiography}

\begin{IEEEbiography}[{\includegraphics[width=1in,clip,keepaspectratio]{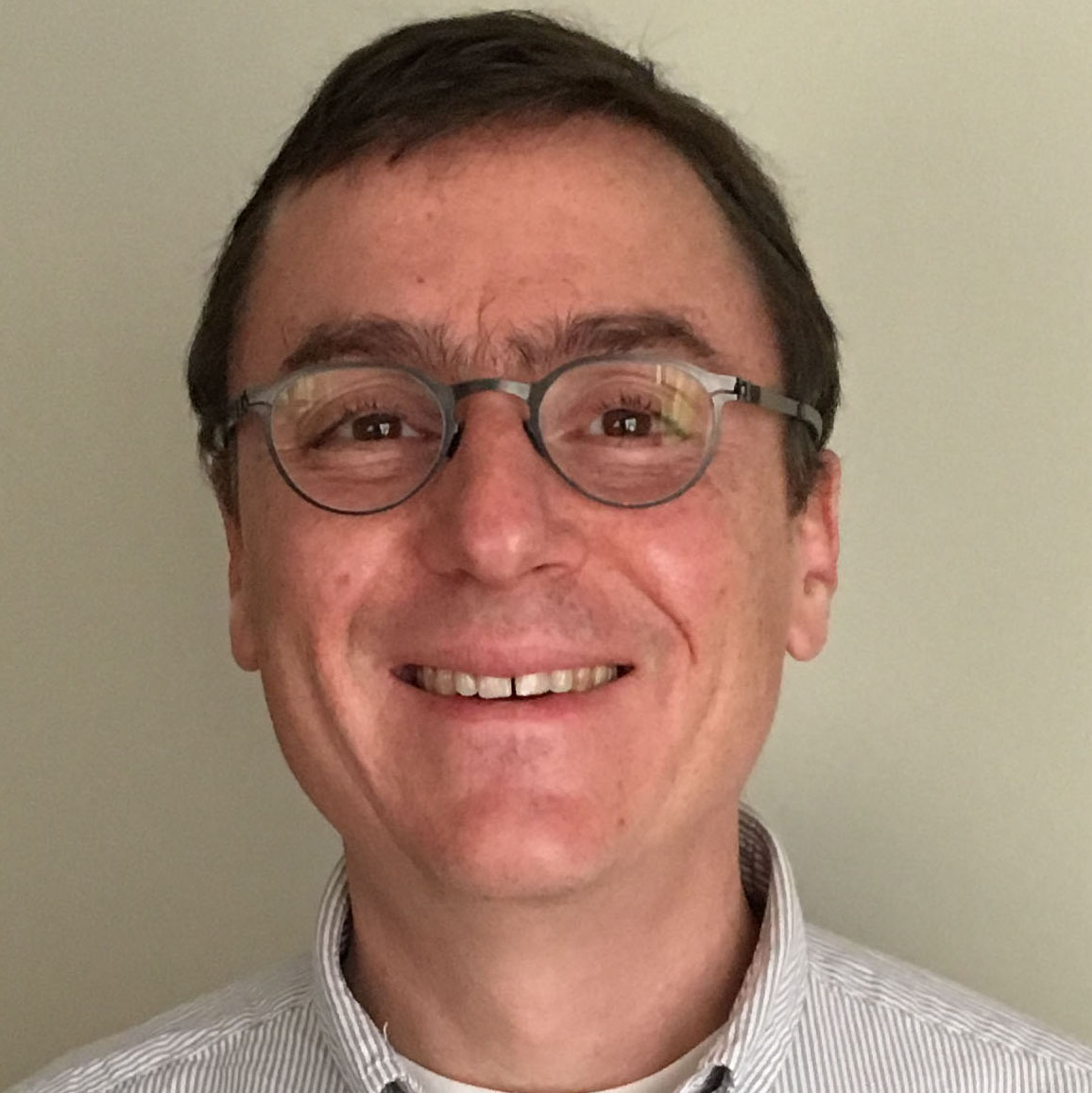}}]{Michael Gleicher}
Michael Gleicher is a Professor in the Department of Computer Sciences at the University of Wisconsin, Madison.  Prof. Gleicher is founder of the Department's Visual Computing Group. He co-directs both the Visual Computing Laboratory and the Collaborative Robotics Laboratory at UW-Madison. His research interests span the range of visual computing, including data visualization, robotics, and virtual reality. 
%He is an expert at developing novel methods for addressing challenges in helping people interact with complex systems. His work has been supported by the National Science Foundation, National Instututes of Health, Mellon Foundation, DARPA, and others. 
Prof. Gleicher was a researcher at The Autodesk Vision Technology Center and in Apple Computer's Advanced Technology Group. He earned his Ph. D. in Computer Science from Carnegie Mellon University, and holds a B.S.E. in Electrical Engineering from Duke University. In 2013-2014, he was a visiting researcher at INRIA Rhone-Alpes. Prof. Gleicher is an ACM Distinguished Scientist.
\end{IEEEbiography}

\end{document}